\documentclass[twocolumn,superscriptaddress,amsmath,amssymb,floatfix,aps,showpacs]{revtex4-1}

\usepackage{amsmath}
\usepackage{graphicx}
\usepackage{hyperref}

\usepackage{color}

\newcommand{\e}[1]{\mathrm{e}^{#1}}

\newcommand{\dir}[2]{\frac{\text{d}#1}{\text{d}#2}}

\newcommand{\x}{\mathrm{X}}
\newcommand{\xs}{\mathrm{X}^*}

\newcommand{\f}[1][]{\mathrm{F}_{#1}}
\newcommand{\fs}[1][]{\mathrm{F}^*_{#1}}

\newcommand{\g}[1][]{\Delta G_{#1}}

\newcommand{\m}[1][]{\mathrm{M}_{#1}}

\newcommand{\kb}{k_{\rm B}}

\newcommand{\teo}[1]{{\color{black} #1}}

\begin{document}

\author{Rory A. Brittain}
\author{Nick S. Jones}
\affiliation{Department of Mathematics, Imperial College London, London, SW7 2AZ, UK}
\author{Thomas E. Ouldridge}
\email{t.ouldridge@imperial.ac.uk}
\affiliation{Imperial College Centre for Synthetic Biology and Department of Bioengineering, Imperial College London, London, SW7 2AZ, UK}

\title{Biochemical Szilard engines for memory-limited inference}

\begin{abstract}
By designing and leveraging an explicit molecular realisation of a measurement-and-feedback-powered Szilard  engine, we investigate the extraction of work from complex environments by minimal machines with finite capacity for memory and decision-making. Living systems perform inference to exploit complex structure, or correlations, in their environment, but the physical limits and underlying cost/benefit trade-offs involved in doing so remain unclear. To probe these questions, we consider a minimal model for a structured environment---a correlated sequence of molecules---and explore mechanisms based on extended Szilard engines for extracting the work stored in these non-equilibrium correlations. We consider systems limited to a single bit of memory making binary `choices' at each step.  We demonstrate that increasingly complex environments allow increasingly sophisticated inference strategies to extract more free energy than simpler alternatives, and argue that optimal design of such machines should also consider the free energy reserves required to ensure robustness against fluctuations due to mistakes. 
\end{abstract}

\maketitle

\section{Introduction}
Living and human-made systems exploit out-of-equilibrium fuel supplies to do useful work. For example, if glucose is present in the environment in higher than equilibrium concentrations relative to carbon dioxide and water, bacteria can power themselves through respiration. Similarly, internal combustion engines use an out of equilibrium concentration of their fuel, i.e. petrol, and are powered by the conversion of fuel and oxygen to carbon dioxide and water.

The amount of work that can be done using the fuel is bounded by the non-equilibrium free energy of the fuel \cite{parrondo2015thermodynamics}. This free energy contains both energetic and entropic terms. As one would expect, if the fuel contains more energy, then, in general, the amount of work that can be done is higher. However, fuels are also more useful if they are in well-defined initial states, with limited microscopic uncertainty. This uncertainty is quantified by the entropy, which is why the entropy contributes to the free energy.

The idea of using high energy fuel is intuitive. If the fuel initially has greater energy than at equilibrium, then that extra energy can be transferred to somewhere else to do useful work as the fuel equilibrates. It is less obvious how to exploit low entropy fuel---nonetheless, entropy is an important component of the free energy stored in biochemical fuel molecules and cellular membrane potentials. For example, the free energy released by converting an ATP molecule to an ADP molecule in a cell is approximately 1.5 times the standard free energy difference between an ATP and an ADP molecule \cite{alberty1992standard}.

Spurred by a desire to understand the fundamental physics of computation and information processing, there has been significant recent interest in the exploitation of purely entropic resources  \cite{mandal2012work,barato2013autonomous,boyd2016identifying,boyd2017correlation,boyd2017leveraging,mcgrath2017biochemical,Stopnitzky2018Physical}. Data arrays are physical systems, and the Shannon entropy of the data contributes to the overall physical entropy of the system. The data itself is therefore a potential resource, and manipulating data has thermodynamic consequences due to changes in the entropy of the data array \cite{landauer1961irreversibility}. 
A data array can have a simple statistical bias towards 1 or 0, and several authors have discussed how such a bias, which implies a low entropy register, might be exploited to perform work \cite{mandal2012work,barato2013autonomous}. A more subtle and equally fundamental possibility is exploiting structure across multiple bits in the array---its entropy can be low due to correlations within the data, rather than an overall bias at the level of individual bits \cite{boyd2016identifying,boyd2017correlation,boyd2017leveraging,mcgrath2017biochemical,Stopnitzky2018Physical}. However, the principles of designing devices to optimally exploit correlations in general settings remain unclear \cite{Stopnitzky2018Physical}.

Although inspired by the physics of computation, the question of how to exploit correlations is also of fundamental biological relevance. If organisms existed in a homogeneous non-equilibrium environment, there would be no need to develop sophisticated information-processing machinery to survive. However, from the chemotaxis system of E. coli to the brains of humans, complex molecular and cellular networks have been evolved to exploit the fact that the environment exhibits correlated fluctuations. These systems rely on the fact that what is sensed at a certain point in space and time contains information about nearby points \cite{mcgrath2017biochemical}. They have evolved even though they are costly to maintain, and despite the fact that the information obtained is limited by features such as the memory and processing power available \cite{govern2014optimal}. However, the fundamental trade-offs that determine the sophistication of these systems are not fully explored.  

In this paper we take steps towards unifying these two perspectives on the exploitation of correlations. We first present a molecular design for a measurement-and-feedback device (a Szilard engine \cite{szilard1929entropieverminderung})  in which the mechanics of the feedback is explicit within the molecular system. We then leverage this construct to propose biomolecular machines that make repeated binary choices about how to act based on measurements of their environment (an array of `molecular bits'). These machines use their single bit of memory to extract chemical work from correlated arrays, demonstrating that it is possible to design minimal biophysical systems that exploit minimal structured environments.

No memory at all is needed to extract all of the available work from an input consisting of an array of uncorrelated subsystems, and simple schemes with one-bit memories can extract all of the stored free energy from Markovian environments. If we increase the complexity of the environment further, by making it a hidden Markov process, 100\% efficiency becomes impossible with a single-bit memory and some implicit inference of the hidden state is required. In this setting, schemes that perform batch averaging to obtain a better estimate of the hidden Markov state can become more efficient than the most direct approaches, at the expense of increased biochemical complexity. We are thus able to construct a minimal thermodynamic setting in which increasingly complex information-processing machinery becomes advantageous in increasingly complex environments.

We first, in section \ref{sec:freeenergy}, give the relevant assumptions and underlying statistical mechanics. Then, in section \ref{sec:modelsystems}, we discuss the previous work on information-exploiting systems and introduce our own model. Numerical methods are briefly discussed in section \ref{sec:numerics}. In section \ref{sec:extract} we demonstrate how work can be extracted from a single molecule in a non-equilibrium state by our setup. Next, in section \ref{sec:szilard}, we discuss how to make a biochemical version of the Szilard engine, which forms the basis of our machines to extract work from correlations. Subsequently we find the maximum amount of work that a device with a persistent memory can extract from a series of correlated bits (section \ref{sec:bounds}). We discuss a device based on the biomolecular Szilard engine that reaches this limit and can extract all of the work available from a Markovian input  in section \ref{sec:markkovianinput}. In sections \ref{sec:nonmarkkovianinput} and \ref{sec:markovmachinehmmwork}, we discuss the limitations of this machine when acting on an input produced by a hidden Markov model. We propose a different machine, in section \ref{sec:batchmachine}, that averages over a batch of multiple input molecules that can extract more work in some cases. Finally, in section \ref{sec:robustness}, we discuss the robustness of such devices to fluctuations in the input.

\section{Materials and Methods}
\subsection{Non-equilibrium generalised free energies and information as a resource}
\label{sec:freeenergy}
In this paper, all physical systems are assumed to be well-described by discrete macrostates of molecules in dilute solution.  Each of these states has an associated chemical free energy, and all systems are in contact with a single heat bath at temperature $T$ \cite{ouldridge2018importance}. We are concerned with small, fluctuating systems, so the state is characterised by a random variable $X$. For any probability distribution over the states of the system $P(X=x)=p(x)$, there is an expected chemical free energy
\begin{equation}
	\langle E(X)\rangle =\sum_xp(x)E(x).
\end{equation}
\teo{Here, $E(x)$ is the chemical free energy of the macrostate $x$, incorporating both the typical energy of $x$,  and any entropic contribution from microscopic variability within $x$ \cite{ouldridge2018importance,ouldridge2018power}. We use $E(x)$ because the chemical free energy plays the same role for macrostates as the energy for fully-resolved microstates.
The distribution $p(x)$ implies an uncertainty in the macrostate $x$, quantified by the  Shannon entropy macrostate $x$ (in nats):}
\begin{equation}
	H(X)= -\sum_xp(x)\ln p(x).
\end{equation}
The generalised non-equilibrium free energy of the system is \cite{esposito2011second}
\begin{equation}
	\mathcal{F}(X)=\langle E(X)\rangle-\kb TH(X).
\end{equation}
The generalised free energy is minimised by the equilibrium distribution to which the system eventually converges.

Now consider a system consisting of two subsystems; the  overall state of the system is the joint random variable $(X,Y)$ where $X$ and $Y$ are the random variables that describe the  individual subsystems. If we assume that the subsystems are not energetically coupled, so that it is possible to write the energy of any joint state as the sum of the energy of the states of the subsystems, then the free energy can be written \cite{parrondo2015thermodynamics}
\begin{equation}
	\mathcal{F}_{\rm joint}(X,Y)=\mathcal{F}_X(X)+\mathcal{F}_Y(Y)+\mathcal{I}(X;Y),
    \label{eq:informationisfreeenergy}
\end{equation}
where $I(X;Y)$ is non-negative the mutual information between the two random variables:
\begin{equation}
	\mathcal{I}(X;Y)=\sum_{x,y}p(x,y)\ln\frac{p(x,y)}{p(x)p(y)}.
\label{eq:MI}
\end{equation}
The mutual information is a measure of how much knowledge of the state of one random variable reduces uncertainty about the state of the other random variable \cite{Elements_of_Information_Theory}.

Eq.~\ref{eq:informationisfreeenergy} shows that there is a real contribution of information to the free energy of a physical system. Fundamentally, correlation between two non-interacting subsystems means that the uncertainty in the state of the joint system is low without a compensating reduction in the energy---work is therefore available.

In terms of the non-equilibrium free energy, the second law of thermodynamics states that the free energy of an isolated system $Z$ can never increase \cite{esposito2011second}:
\begin{equation}
	\Delta\mathcal{F}(Z)=\mathcal{F}(Z,t+\tau)-\mathcal{F}(Z,t)\leq0.
\end{equation}
Let $Z$ consist of two non-interacting subsystems $X$ and $Y$, as in equation \ref{eq:informationisfreeenergy}, and assume the mutual information between subsystems is zero at the time $t$. Then for any process between time $t$ and $t+\tau$ that leaves $X$ and $Y$ non-interacting in the final state,
\begin{equation}
	\Delta\mathcal{F}_Y(Y)\leq-\Delta\mathcal{F}_X(X).
\end{equation}
The reduction in the free energy of $X$ can be used to increase the free energy of $Y$ by an amount up to the magnitude of the change in free energy of $X$. In this paper we will refer to this increase of free energy of $Y$ as `work' being performed on the physical system, with work being a shorthand for the more formal term `chemical work' \cite{van2015ensemble,ouldridge2018power}. Therefore, in a process that reduces the free energy of a subsystem $X$ by $\Delta\mathcal{F}_X$ a work of $W\geq-\Delta\mathcal{F}_X$ can be done on another subsystem.

\subsection{Model systems}
\label{sec:modelsystems}

\subsubsection{Prior models}
In this work we consider machines designed to extract work from a non-equilibrium series of bits with both the machines and the bits rendered as biomolecules. These devices exploit pre-existing information within the input via a series of measurement and feedback operations implemented through a 1-bit memory.
 We now summarise prior work on
\teo{theoretical constructs for the exploitation of information  to put this study into context.

Underpinning our device is an exact and explicit biochemical formulation of the Szilard engine \cite{szilard1929entropieverminderung}. Szilard used this thought experiment to argue against the possibility of an observer violating the second law by measuring a system's equilibrium fluctuations and subsequently using feedback to exploit them -- a problem originally considered by James Clerk Maxwell \cite{dougal2016kelvin} in the context of his infamous `Demon'. Szilard explained that any exploitation required an `ominous coupling' between the measured system and the system that performs the feedback---a correlation that persists beyond the physical decoupling of the two degrees of freedom.  
He argued that such a `measurement' cannot be performed without a `compensation' that preserves the second law. Eq.~\ref{eq:informationisfreeenergy} is a more modern formulation of this argument: correlations between decoupled degrees of freedom store free energy, and therefore producing them has a thermodynamic cost. In his original work,  Szilard analysed explicit mechanisms for both the measurement and exploitation separately, but he did not analyse a full cycle of measurement and feedback in a single system. Furthermore, he did not consider the challenge of extracting work from a series of correlated inputs. 

Recent advances in nonequilibrium thermodynamics have prompted a resurgence of interest in Maxwell Demons, Szilard Engines and related systems. One significant avenue of investigation has focused on bipartite systems, in which two subsystems are physically coupled but undergo individual transitions. It has been shown that the full second law of bipartite systems can be decomposed into individual second laws for each subsystem \cite{horowitz2014thermodynamics,barato2014efficiency,allahverdyan2009thermodynamic}. These individual second laws contain an additional term describing how transitions within the subsystems influence the information shared between them. If this  term has the right sign, it can allow the other contributions to the entropy production in one of the subsystems to be negative---an apparent violation of the second law for an observer that is aware of only one subsystem. Esposito and collaborators have shown, through experiment and theory, that this effect can be observed even when the net energy transfer between the two subsystems is zero \cite{strasberg2013thermodynamics,koski2015on-chip}, describing such systems as `true Maxwell Demons'. 
However, these devices don't demonstrate the kind of behaviour seen in Szilard's Engine or Maxwell's Demon in the sense of storing, then subsequently exploiting, free energy within correlations between non-interacting systems.

Simultaneously, a second major class of systems has arisen as a testbed for ideas about the thermodynamics of information: machines designed to extract work from a non-equilibrium series of bits. }
The first detailed analysis of such a machine was performed by Mandal and Jarzynski \cite{mandal2012work}. The authors considered a three state device that couples to each bit in an input sequence for a period of time before being moved to the next bit. The machine changes state stochastically and couples the changing of the state of the input bit to the raising and lowering of a mass in a gravitational field. Although the authors pointed out that correlations within the tape could store free energy, their actual design could only exploit the overall bias of the input bits towards either 0 or 1. The device is powered by an increase in the entropy of its input, rather than a change in its energy, but the fundamental principle is not dissimilar to a device that exploits the difference in pressure between two volumes of ideal gasses, which is also entropic in nature. The analogy is particularly vivid if one assigns a `0' to gas particles arriving from the left of a piston, and `1' to particles arriving from the right.

This model was extended to allow the device to step stochastically along its tape, and furnished with a chemical realisation, by Barato and Seifert \cite{barato2013autonomous}. In neither case is information in the environment---in the sense of structure induced by correlations---exploited, and there is no feedback from the state of the tape to the operation of the device.


Horowitz {\it et al.} discussed a device that interacts with a series of two-state systems via a process of measurement and feedback \cite{horowitz2013imitating}. The input was an equilibrium system, however, without correlations between successive subsystems. Hence the mechanism of measurement and feedback, which was \teo{not explicitly described as an inherent part of the system under study}, must necessarily consume at least as much work as could be extracted in the exploitation step. \teo{Diana {\it et al.} considered the converse problem: using  measurement and feedback to reduce the work required to set an array of bits to 0 \cite{diana2013finite-time}. Again, however, correlations within the tape were not considered, and the feedback mechanism was implicit.}

Boyd {\it et al.} have sought to develop machines that extract work from `temporal' correlations between successive bits \cite{boyd2016identifying,boyd2017correlation,boyd2017leveraging}. The authors consider, in a similar fashion to the previous models, a machine with a number of discrete states coupled to successive bits in a long string of inputs. These machines are intended to extract work from tapes that have no overall bias towards one state or the other but contain correlations between the state of  bits. 
As has been highlighted by Stopnitzky {\it et al.}, however, the machines in these works were designed without `reversibly embeddable' dynamics---a necessity if the machines are to operate without external control, as was assumed \cite{Stopnitzky2018Physical}. Stopnitzky {\it et al.} did present systems with reversibly embeddable dynamics that extract positive work from a perfect sequence of alternating 1s and 0s, but the efficiency was very low. The extraction of work from perfectly correlated systems has also been analysed in a quantum mechanical setting \cite{chapman2015autonomous}. 

A biochemical machine for exploiting correlated pairs of molecules was presented by McGrath et al.  \cite{mcgrath2017biochemical}. Although information between non-interacting molecules is indeed exploited in this work, the nature of the correlations---which are much more simple than those in a string of bits encountered one after the other---allow a particularly straightforward, memory-free approach. In effect, the pairs of molecules could be described as a single 4-state non-equilibrium system, and processed in isolation from other pairs. 

The lack of concrete physical rendering in some of these models \cite{mandal2012work,diana2013finite-time,boyd2016identifying,boyd2017correlation,boyd2017leveraging,Stopnitzky2018Physical,chapman2015autonomous}
makes the machines mysterious and increases the scope for error, as discussed in \cite{Stopnitzky2018Physical}. If the inputs are simply described as an abstract string of bits without any explanation of their physical instantiation, the low entropy of the data is thereby made to seem like a new, and almost non-physical source of work. Measurement-and-feedback-driven devices in which the feedback mechanism is implicit can also ignore some of the costs of the process; down-play the challenges of inducing feedback-driven behaviour \teo{in which one component first influences the evolution of the other, and then {\it vice versa}}. For those unfamiliar with the field, such an approach can provide misleading intuition as to how the measurement must be stored, as we will discuss. 

\subsubsection{Molecular implementation of a measurement-and-feedback machine} 
\label{sec:our_model}
\begin{figure}[!]
	\includegraphics[width=\linewidth]{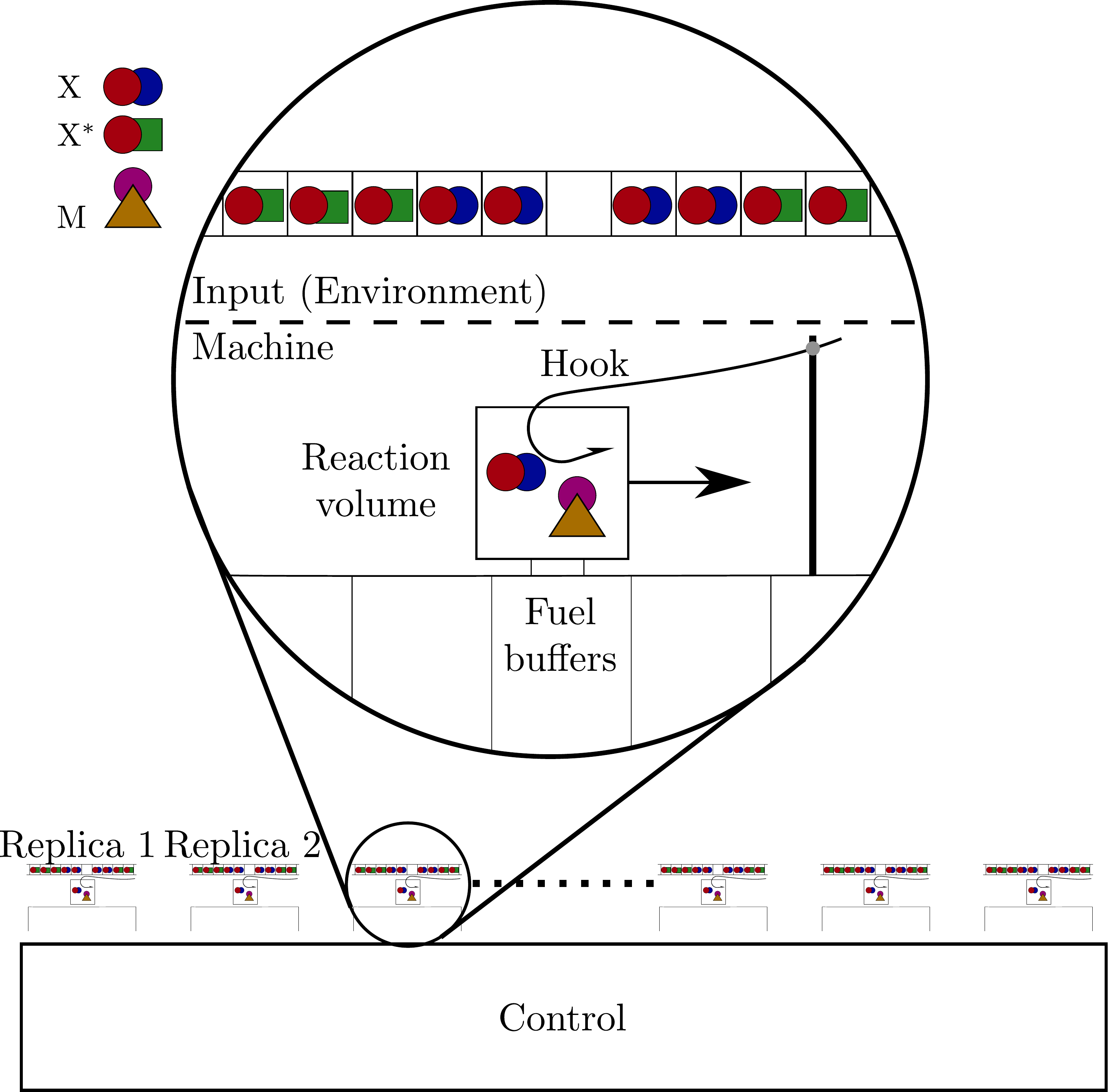}
    \caption{\textit{General design of model systems considered in this work.} A series of input molecules each contained in a box (each one being an analogue of Szilard's box containing a gas molecule); a reaction volume; a series of buffers (acting as work reservoirs like the weight in the conventional Szilard engine); and a hook. One input molecule is moved from its box to the reaction volume by the hook. The reaction volume is then connected to a series of buffers in succession containing different concentrations of fuel molecules. \teo{These buffers exchange molecules with the reaction volume, driving the reaction in one direction or another. The net transfer of fuel to buffers of high chemical potential  corresponds to the extraction of work from the input molecule.} Then the hook moves the input molecule from the reaction volume back to its original box, and the next input molecule can be moved to the reaction volume. The dashed line separates the input/environment from the machine. The control can operate multiple replicas of the system simultaneously.}
    \label{fig:setup}
\end{figure}
We now present a general description of the devices considered in this work. \teo{Although our devices would be challenging to engineer, operate in ideal limits, and are not direct models of living systems, we nonetheless render the machines, and the input bits, as biomolecules. All operations, including the measurement and feedback, are driven by a concrete molecular mechanisms that are explicitly part of the devices themselves. By considering a concrete realisation, even in an idealised limit, we can explore the limits of what is thermodynamically possible in a positive sense, rather than simply exploring the space of systems that are not forbidden by a particular aspect of the second law \cite{ouldridge2018power}. Furthermore, we demonstrate the true complexity of systems required to instantiate efficient measurement-and-feedback systems like Szilard engines.}

The schematic set-up of our devices is shown in figure \ref{fig:setup}. The model consists of an input, a reaction volume, a series of chemical buffers, and a molecular `hook' that can bind to the input molecules independently of their state \cite{ouldridge2017thermodynamics,ouldridge2017fundamental}. The input is a series of small boxes each containing a single input molecule. This molecule can be in one of two strongly metastable states, $\x$ and $\xs$ so these input molecules represent a string of bits. This input is a minimal analogue of a fluctuating chemical environment, as experienced by single-celled organisms \cite{micali2016bacterial,becker2015optimal,parkinson2015signaling,mitchell2009adaptive}.

The rest of the system is our machine, a minimal analogue of an organism exploiting its environment. The machine functions by transferring molecules to and from its reaction volume via the molecular hook. Once in the reaction volume, input molecules undergo reactions with molecules that are internal to the system---for example, a molecule M encoding the memory. These reactions are coupled to large fuel buffers that collectively allow the machine to store the work extracted from the environment, similar to Refs. \cite{barato2013autonomous,mcgrath2017biochemical}. The buffers are the molecular analogue of a weight in a gravitational field that can be lifted by the system \cite{mandal2012work,boyd2016identifying,boyd2017correlation,boyd2017leveraging}. \teo{To perform this role, the buffers should be sufficiently large that any reactions have a negligible effect on the probability distribution of bath macrostates \cite{ouldridge2018power}; in this limit, the buffer state is purely a deterministic concentration, whereas the input and other molecules within the reaction volume are represented through stochastic variables describing the fluctuating chemical macrostate. Recent experimental work from Joesaar {\it et al.} has demonstrated how molecules encapsulated within `proteinosome' reaction volumes can be coupled to time-varying external buffer conditions, with the buffer molecules able to diffuse in and out of the reaction volumes and participate in reactions with the encapsulated species, as we require \cite{joesaar2019dna-based}}. 

Details of how a molecular hook might operate are given in appendix \ref{sec:apphook}. Such a mechanism can transfer molecules to and from the reaction volume with no net expenditure of work, provided that the hook is controlled by a particular quasistatic protocol. \teo{The hook thus represents a work-free mechanism of ingesting and excreting molecules in a controlled manner. An alternative model without ingestion would have input molecules attached to consecutive sites on a polymer tape \cite{mcgrath2017biochemical}; the machine would then interact with one or more of these molecules at any one time, based on proximity.}

In this work we shall assume that buffer concentrations are manipulated  by a well-defined protocol \cite{ouldridge2017thermodynamics,ouldridge2018importance,rao2016nonequilibrium,schmiedl2007stochastic}, as illustrated in figure \ref{fig:exploit_example}. 
 These  protocols will not follow directly from the dynamics of degrees of freedom explicitly modelled---they are essentially externally imposed. Our system is then non-autonomous. 
\teo{We use externally applied protocols for two reasons. Firstly, because it allows us to design mechanisms in which first measurement of $Y$ by $X$, and then feedback to exploit $Y$ using $X$, are performed sequentially, as in Szilard's engine (see Section~III A); and secondly, because driving forces can be increased in a quasistatic  manner (see Section~II D). Both features are essential if we are to maximise both the efficiency of the reactions, and the reliability of the implemented information-processing strategy: quasistatic manipulation allows us to push reactions implementing each step of the process to 100\% completion, efficiently. We can then focus purely on the constraints on work extraction that arise from the fact that the implementable information-processing strategies are limited by the finite size of the device's memory.}

Crucially, however, although our systems require external protocols, the protocols themselves require no decision-making intelligence; the same series of manipulations will be applied repeatedly, without feedback from the state of the system. All `decisions' and feedback strategies must be made by the molecules that are explicitly represented.
By avoiding protocols that require external decision-making dependent on the state of the system, we avoid implicit costs that have caused much of the confusion in the thermodynamics of computation, since the original thought experiment of Maxwell \cite{maxwell1891theory}. In principle, the protocols we invoke could be applied in parallel to an arbitrarily large number of replicas (as shown in figure \ref{fig:setup}), rendering the marginal cost per machine of the external protocol negligible. Indeed, this is the assumption usually made with macroscopic thermodynamics. By contrast, if separate decisions had to be made for each replica, the economy of scale would not exist. Although the use of external control makes the individual devices a weaker analogy for single, autonomous organisms, the combined set-up of many devices and their controller is then an analogy for a single, albeit more complex, organism. We note in passing that the need for a quasistatic protocol to control the hook is equivalent to the need for a quasistatic protocol to deterministically advance the tape with which a machine interacts, as has previously been assumed in many bit-driven machines \cite{mandal2012work,boyd2016identifying,boyd2017correlation,boyd2017leveraging,Stopnitzky2018Physical}---our physical instantiation makes this need clearer.


\subsection{Numerics}
\label{sec:numerics}
\teo{All devices studied in this work produce a deterministic output of extracted work given a specific input sequence. Numerical results presented in this work are therefore obtained either by exhaustive summation of short input strings, or by sampling of long input strings by simulating the underlying generative model.  The code and data to produce the figures in this paper can be found at \url{https://doi.org/10.5281/zenodo.1976932}.}

\subsection{Example system and calculation: Extracting work from a biased environment}
\label{sec:extract}
\begin{figure}[!]
	\includegraphics[width=\linewidth]{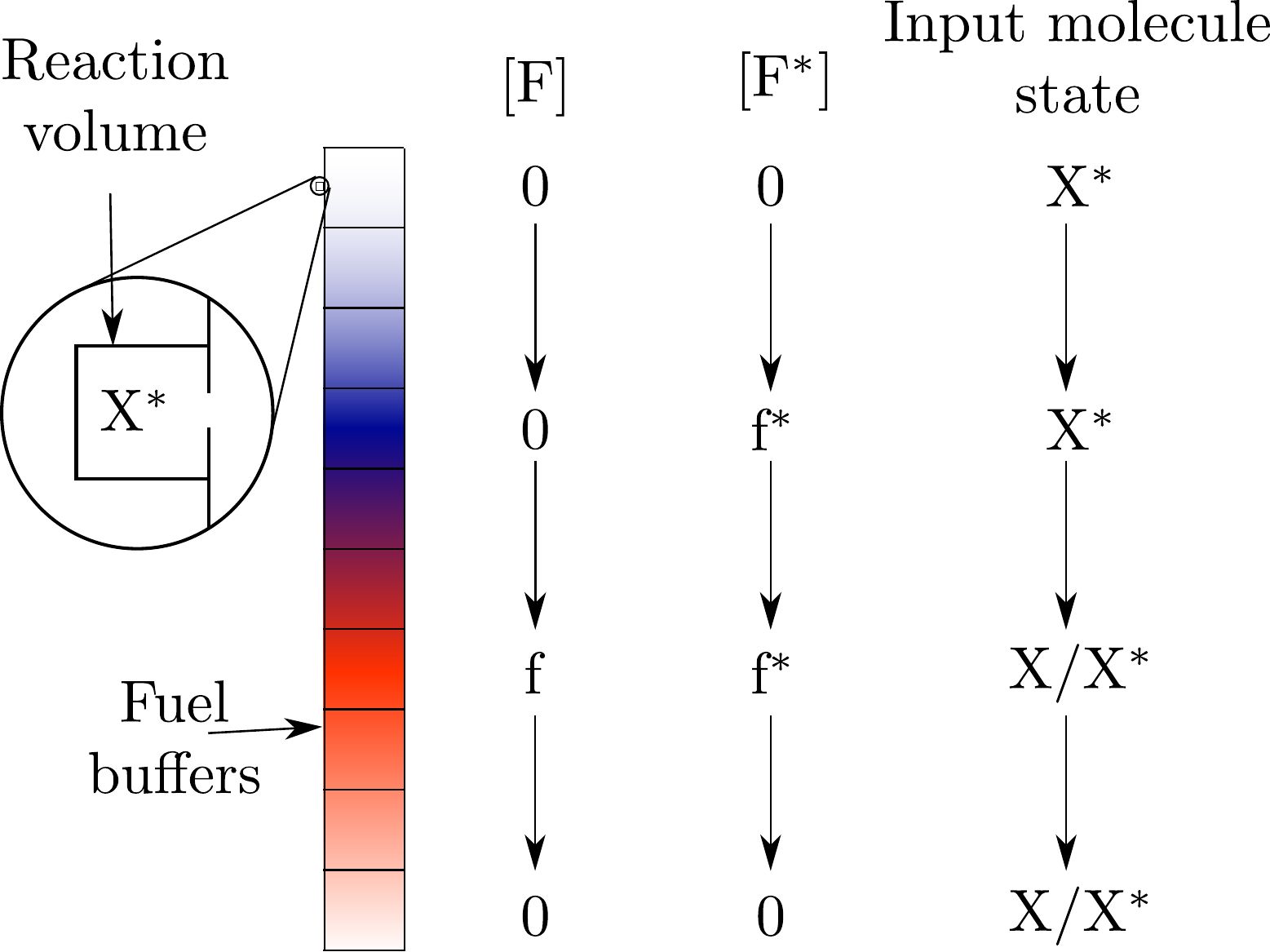}
    \caption{\textit{ Fuel manipulation protocol to extract maximal work for a known input state.} The concentration of the fuel molecules in the reaction volume is set by connecting the reaction volume with a chemical buffer. The concentrations can be gradually changed by connecting the reaction volume to a series of buffers with a small change in concentration between adjacent buffers. \teo{In this case, no reactions are initially possible since both fuel concentrations are zero. $[\fs]$ is first increased so that the presence of $\xs$ and the absence of $\x$ is consistent with the equilibrium implied by the imbalance in $[\fs]/[\f]$. $[\f]$ is then increased, allowing $\xs$ to convert to $\x$ in a quasistatic manner and extracting work from the process, as molecules of $\f$ at low chemical potential are converted to molecules of $\fs$ at high chemical potential. Finally, both fuels are taken to zero concentration. }}
    \label{fig:exploit_example}
\end{figure}
We illustrate the operation and analysis of the set-up outlined in section \ref{sec:our_model} by demonstrating it in the simplest possible context. We consider the reversible extraction of work from a low entropy input by increasing the free energy of a chemical buffer. In this setting, the input array consists of input molecules each initially in the state $\xs$ with 100\% probability. The $\x$ and $\xs$ states of the input molecules have equal intrinsic free energy so in equilibrium a single input molecule is equally likely to be in either state. Therefore, it is possible to extract a work of $\kb T\ln2$ per input molecule from the environment. 

Each input molecule is transferred to and from the reaction volume by a hook with no net work expenditure, as outlined in appendix \ref{sec:apphook}. When the input is in the reaction volume, we extract work by increasing the free energy of a bath of fuel molecules $\f$ and $\fs$ with chemical potentials $\mu_{\f}$ and $\mu_{\fs}$. To do this we need a chemical reaction
\begin{equation}
	\x+\fs\rightleftharpoons\xs+\f,
    \label{eq:conversion}
\end{equation}
which couples the interconversion of $\x$ and $\xs$ to the interconversion of $\f$ and $\fs$. The interconversion of $\x$ and $\xs$, or $\f$ and $\fs$, is assumed to be infinitely slow except via this reaction. No other molecules, such as those representing a memory, are necessary in this simple context. \teo{The central idea is that an excess of $\xs$ can be used to pump $\f$ into $\fs$ against a chemical potential difference, storing work in the buffers just as traditional heat engines store work by lifting a weight. We now consider the details of how this work might be done}.

It is possible to extract some work by connecting the $\xs$ molecule to  a single bath of $\f$ and $\fs$ molecules with a high concentration of $\fs$, so that $\mu_{\fs} > \mu_{\f}$. Both the input and the bath are individually out of equilibrium, and tend to drive the reaction in Eq.~\ref{eq:conversion} in opposite directions. In this case, the drive from the input is stronger and the reaction in Eq.~\ref{eq:conversion} proceeds from right to left, with the input doing work on the bath.  Over time, the bias of the input will decrease until the driving force of both contributions cancel; although the bath and the input are individually still out of equilibrium and store free energy, the input has reach a bias which is in equilibrium with driving force of the  bath. At this point, the input will be in state $\x$ with probability $1/(1+\e{-\beta\g})$ and in state $\xs$ with probability $\e{-\beta\g}/(1+\e{-\beta\g})$ where $\beta=1/(\kb T)$, $\g=\mu_{\f}-\mu_{\fs} <0$. During this relaxation to equilibrium,  $1/(1+\e{-\beta\g})$ molecules of $\f$ are converted to $\fs$ on average. Therefore, the free energy of the bath is changed by $-\g/(1+\e{-\beta\g})$---this is the work extracted per input molecule.

Different choices of $\g$ lead to different values of the work; however, \teo{$-\g/(1+\e{-\beta\g})$ has a maximum of} $\approx0.28\kb T$, which is less than $\kb T\ln2$. This protocol has not extracted all of the work available; indeed the input molecule has not even reached its equilibrium distribution, so it is still a store of free energy. Thus the input molecule could be put in contact with a second bath with a lower concentration of $\fs$ molecules but still with an excess of $\fs$ above the equilibrium concentration and some more work could be extracted.

If the input molecule is connected to two successive baths with a non-infinitesimal difference in fuel concentrations, then the input molecule undergoes a thermodynamically irreversible relaxation, with some fraction of the free energy being wasted. However, if we take this idea of connecting the input molecule to successive baths with lower $\Delta G= \mu_{\f}-\mu_{\fs}$ to the limit of a continuous change in $\Delta G$ we get a quasistatic process with no irreversible relaxations to equilibrium: the system is at equilibrium with the bath(s) at all points in time. This protocol is achieved by connecting the reaction volume to a large number of baths in succession for enough time to reach equilibrium with each bath as shown in figure \ref{fig:exploit_example}. There is only a small change in concentration of fuel molecules between successive baths. Therefore, in the limit of infinite baths and infinitesimal changes in concentration the reaction volume experiences a quasistatic change in the concentrations of the fuel molecules.

The specific protocol of fuel molecule concentrations, illustrated in figure \ref{fig:exploit_example}, is as follows. Initially $[\f]=[\fs]=0$ so reaction \ref{eq:conversion} cannot occur. Then $[\fs]$ is slowly increased up to an appreciable value we name $f^*$. The reason the concentration must be increased slowly is so that fuel molecules are not irreversibly transferred between different buffers via the reaction volume. The reaction in equation \ref{eq:conversion} still cannot occur, since only $\xs$ and $\fs$ are present. Then, $\f$ is slowly increased. Now reaction \ref{eq:conversion} can occur; although, initially, the rate of converting $\xs$ to $\x$ is much slower than the reverse so the input molecule is still in state $\xs$ with high probability. $[\f]$ is increased to $f$, which is the concentration at which the free energy change in reaction \ref{eq:conversion} is $\Delta G=0$, so the $\x$ and $\xs$ states are equally likely.

\begin{figure*}[!]
    \centering
	\includegraphics[width=\linewidth]{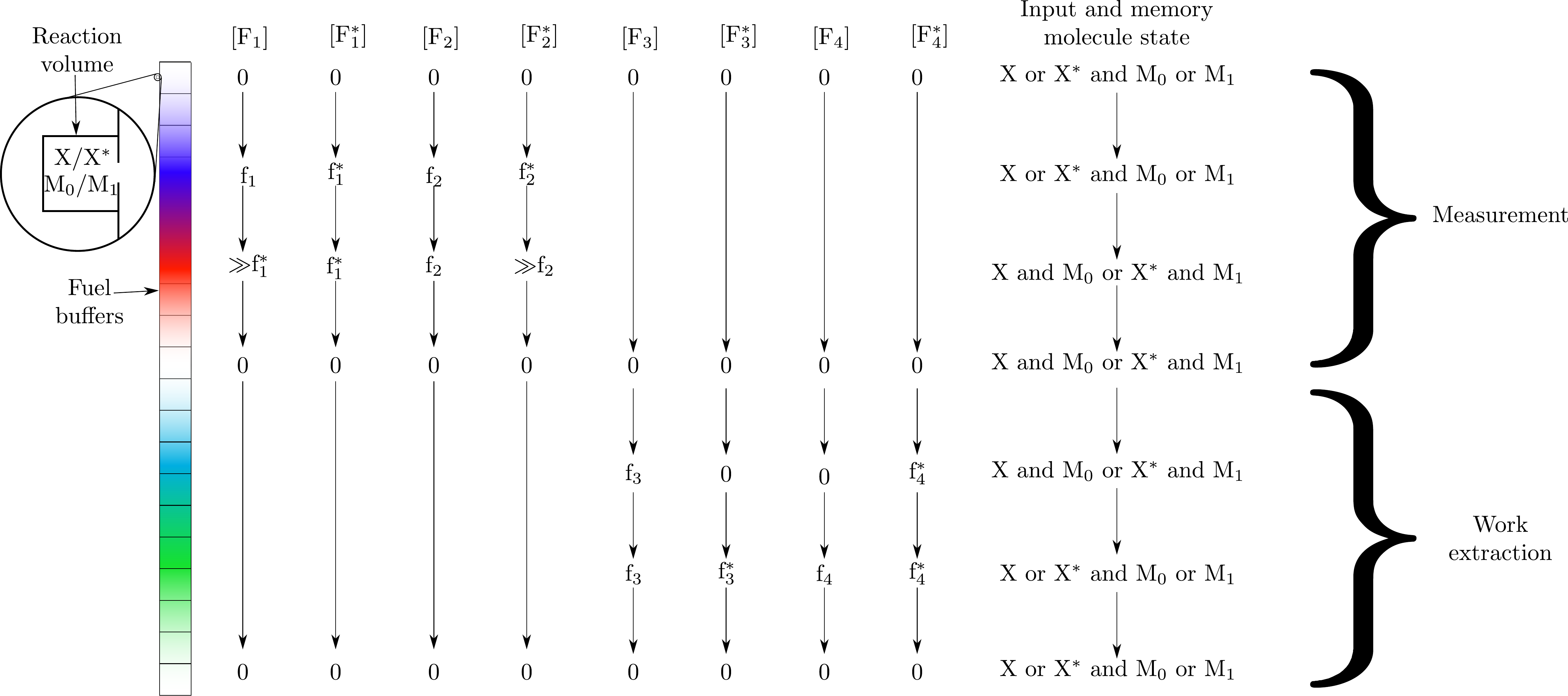}
    \caption{\textit{ Fuel manipulation protocol for the biochemical implementation of the Szilard engine.} In the first stage the concentrations of fuel molecules are changed to set the memory molecule to $\m[0]$ if the input molecule is $\x$ and $\m[1]$ if the input molecule is $\xs$. In the second stage work is extracted from the correlation between the input molecule and the memory molecule. \teo{The ability to first use X to set M via $\f[1]$ and $\f[2]$, and then couple M to the exploitation of X using $\f[3]$ and $\f[4]$, is the key to implementing Szilard's engine.
In both stages, as in figure~\ref{fig:exploit_example}, the fuels are initially set to zero so no reaction can happen. They are then increased to finite values, but maintaining a ratio consistent with the equilibrium implied by the current state of the X and M molecules (right hand column), so that no net reactions happen. From this point, the concentrations of fuels are then manipulated to drive the X and M molecules towards the desired state. Finally, the fuel concentrations are set to zero again. }}
    \label{fig:szilard_protocols}
\end{figure*}


To calculate the average work extracted in this quasistatic process we consider the increase in free energy of the $\f/\fs$ baths. Let the probability of the input molecule occupying state $\xs$, when equilibrated with a buffer with a chemical potential difference of $\g$, be $p_{\g}(\xs)$. A change in chemical potential difference of $\g-\delta\g$ to $\g$ is then associated with a probability change of $p_{\g}(\xs)-p_{\g-\delta\g}(\xs) \approx \dir{p_{\g}(\xs)}{\g}\delta\g$. This change is also equal to the number of $\fs$ molecules that are on average converted to $\f$ molecules when the reaction volume is exposed to a new buffer. Therefore, the free energy of the bath increases by $\g\dir{p_{\g}(\xs)}{\g}\delta\g$ on average.

Taking the limit of infinitely many baths, we integrate the total work done
\begin{align}
	W&=\int_\infty^0\g\dir{p_{\g}(\xs)}{\g}\text{d}\g\nonumber\\
    &=\kb T \ln2,
\end{align}
where we integrate by parts and recall that $p_{\g}(\xs)=\e{-\beta\g}/(1+\e{-\beta\g})$. The quasistatic protocol is therefore able to recover all free energy stored in the initial low entropy state, $\kb T\ln2$, as work. In performing this calculation, we have ignored external costs associated with generating the quasistatic protocol, for reasons outlined in \ref{sec:our_model}. The transfer of molecules between adjacent buffers, mediated by the reaction volume, has a cost that tends to zero as the concentration difference between buffers tends to zero. With the basic approach to set-up and analyse of our machines explained, we can discuss specific measurement and feedback processes.

\section{Results}
\subsection{A Biochemical Szilard engine}
\label{sec:szilard}
Before analysing structured environments, we first present a measurement-and-feedback device that acts on a single binary input.  This simpler setting allows us to illustrate the explicit measurement-and-feedback cycle that will underlie all the devices in this work. In particular, we demonstrate a mechanism by which the input is first able to influence the state of a memory, and subsequently the influence is reversed so that the state of the memory affects how work is extracted from the input.


The biochemical Szilard engine consists of an input molecule, a memory molecule, and chemical fuel buffers that are used to supply or recover chemical work; \teo{to implement a single cycle, we do not require a series of inputs, or a molecular hook}. The input molecule is in one of two states: $\x$ or $\xs$. For simplicity, we assume the states have equivalent intrinsic free energy, and that the system is in equilibrium: the molecule is then found in each state with probability $\frac{1}{2}$. The memory molecule also has two states with equivalent intrinsic free energy, and is initially in state $\m[0]$ with probability $\frac{1}{2}$ and in state $\m[1]$ with probability $\frac{1}{2}$.

To `measure' the state of the input means to set the state of the memory to $\m[0]$ if the input is $\x$ or to $\m[1]$ if the input is $\xs$: we correlate the states. This step follows the optimal copy protocol in \cite{ouldridge2017thermodynamics} and can be done using the chemical reactions
\begin{align}
	\m[0]+\x+\fs[1]&\rightleftharpoons\m[1]+\x+\f[1],\nonumber\\
	\m[0]+\xs+\fs[2]&\rightleftharpoons\m[1]+\xs+\f[2],
    \label{eq:szilardmeasure}
\end{align}
where $\f[1]$, $\fs[1]$, $\f[2]$ and $\fs[2]$ are a fuel molecules that are present in excess, and $\x$ and $\x^*$ act as catalysts for the transformation of $\m$ between its states. Interconversions other than via the catalytic reactions in equation \ref{eq:szilardmeasure} are assumed to be so slow as to be negligible. \teo{The central idea is to drive the catalytic reactions in opposite directions through fuel imbalances, so that M can be set to $\m[0]$ in the presence of $\x$, and to $\m[1]$ in the presence of $\xs$. }

The selective catalysis in equation \ref{eq:szilardmeasure} is an approximation of the behaviour demonstrated by bi-functional kinases  in cell signalling networks \cite{stock2000two}, and can also be engineered from nucleic acid networks (see appendix \ref{sec:appszilarddna} for details). The free energy changes of the reactions and the reaction rates can be controlled by the concentrations of the fuel molecules, as in the simple example in section \ref{sec:extract}.

It would be possible to set the memory molecule $\m$ to the correct state by directly coupling to a buffer with $[\f[1]]\gg[\fs[1]]$ and $[\fs[2]]\gg[\f[2]]$. As in section \ref{sec:extract}, however, the associated process would be thermodynamically irreversible, wasting the ability of the fuel buffer to do useful work. We therefore change the fuel concentrations quasistatically, as illustrated in  figure \ref{fig:szilard_protocols}, gradually forcing the memory to the $\m[0]$ state when in the presence of $\x$, and to the $\m[1]$ state in the presence of $\xs$. 


Initially, $[\f[1]]$, $[\fs[1]]$, $[\f[2]]$ and $[\fs[2]]$ are all set to zero. The reactions in equation \ref{eq:szilardmeasure} therefore cannot occur. Then, the concentrations are simultaneously increased at a fixed ratio of $[\f[1]]/[\fs[1]]$ and $[\f[2]]/[\fs[2]]$ that maintain an overall free energy change of zero for the reactions in equation \ref{eq:szilardmeasure}. One of these interconversions (determined by whether there is an $\x$ or and $\xs$ present) now occurs at an appreciable rate, but forwards reactions exactly balance reverse reactions so there is no overall change in the probability of observation of $\m[0]$ and $\m[1]$. 

Next, $[\f[1]]$ and $[\fs[2]]$ are increased while $[\fs[1]]$ and $[\f[2]]$ are kept constant. As a result, the first reaction in equation~\ref{eq:szilardmeasure} is pushed to the left and the second to the right. Consequently, if the input is $\x$ then the memory molecule is more likely to be $\m[0]$ and if the input is $\xs$ the memory molecule is more likely to be $\m[1]$. Eventually, when $[\f[1]]$ and $[\fs[2]]$ have been increased so that $[\f[1]]\gg[\fs[1]]$ and $[\fs[2]]\gg[\f[2]]$, the memory molecule $\m$ will be perfectly correlated with the input $\x$. Next, $[\f[1]]$, $[\fs[1]]$, $[\f[2]]$ and $[\fs[2]]$ are decreased while maintaining $[\f[1]]\gg[\fs[1]]$ and $[\fs[2]]\gg[\f[2]]$ until $[\fs[1]]=[\f[2]]=0$. Finally $[\f[1]]$ and $[\fs[2]]$ are decreased to zero. Now the reactions in equation \ref{eq:szilardmeasure}, again, cannot occur so the memory molecule is fixed to be $\m[0]$ if the input is $\x$ and $\m[1]$ if the input is $\xs$.

In this correlated state the entropy of the combined $(\x, \m)$ system is $\kb\ln2$ because there are two equally likely states: $(\x, \m)= (\xs, \m[1])$ or $ (\x, \m[0])$. Prior to measurement the entropy was $\kb\ln 4$ because the four combinations of $\x$ and  $\m$  are equally likely. Thus the entropy of the system has decreased by $\kb\ln2$ and so the free energy of the system has increased by $\kb T\ln2$. 

The increase in free energy of $(\x, \m)$ is compensated by a decrease in the free energy of the buffers. This decrease can be calculated as in section \ref{sec:extract}, except with the limits on the integral reversed and considering two equally likely possibilities: either the input molecule was $\x$ and the concentrations of $\f[1]$ and $\fs[1]$ are changed due to the first reaction in equation \ref{eq:szilardmeasure}, or the input molecule was $\xs$ and the concentrations of $\f[2]$ and $\fs[2]$ are changed due to the section reaction in equation \ref{eq:szilardmeasure}. The result is that the free energy change of the buffers is $-\kb T\ln2$, which exactly cancels the free energy increase of $(\x, \m)$, as it should because the process is thermodynamically reversible (see appendix \ref{sec:appszilardmeasure} for more details on this calculation). This reduction in free energy of the buffers is the `cost' to measurement that was recognised by Szilard as the resolution to the Maxwell's demon paradox \cite{szilard1929entropieverminderung}.

We now consider the feedback step. The device extracts chemical work from the correlated state by allowing the input molecule to evolve in a manner that reflects the outcome of the measurement. The machine uses the reactions
\begin{align}
	\m[0]+\x+\fs[3]&\rightleftharpoons\m[0]+\xs+\f[3],\nonumber\\
	\m[1]+\x+\fs[4]&\rightleftharpoons\m[1]+\xs+\f[4],
    \label{eq:szilardextract}
\end{align}
in which the $\fs[3]$, $\fs[4]$, $\f[3]$ and $\f[4]$ are further fuel molecules. Now, $\m[0]$ and $\m[1]$ act as catalysts for the transformation of $\x$ between its states; non-catalysed reactions are again assumed to be impossible. $\m$ and $\x$ must therefore be mutual bifunctional catalysts, which can be effectively switched on and off by modulating fuel concentrations. This explicit rendering demonstrates the complexity necessary in a minimal measurement-and-feedback device such as Szilard's engine, in which the memory and input must reverse their roles as the determinants of the dynamics. A design based on DNA strand displacement \cite{soloveichik2010dna,cardelli2013two,chen2013programmable} is presented in appendix \ref{sec:appszilarddna}.

As in the measurement step, the reaction rates are slowly manipulated by coupling to buffers with different concentrations of fuel molecules. Initially $[\f[3]]=[\fs[3]]=[\f[4]]=[\fs[4]]=0$, along with the fuels used in the measurement process, and no reactions occur. Subsequently, $[\f[3]]$ and $[\fs[4]]$ are increased. At this point the reactions in equation \ref{eq:szilardextract} do not occur since the right combination of fuels and substrates are not present.  
 
Next, $[\fs[3]]$ and $[\f[4]]$ are increased until the free-energy change of reactions in equation \ref{eq:szilardextract} is zero. As a result, the input molecule is slowly decorrelated from the memory, but the memory state determines which fuel buffer the input couples to during this process. Finally, $[\f[3]]$, $[\fs[3]]$, $[\f[4]]$ and $[\fs[4]]$ are decreased to zero while maintaining a free-energy change of zero for the reactions in equation \ref{eq:szilardextract}. 

As with the measurement, we can calculate the change in free energy of the input molecule and measurement molecule system and the chemical work done by the chemical fuel buffers. This extraction step is essentially the reverse of the measurement step, so the free energy of the input molecule and measurement molecule system decreases by $\kb T\ln2$ while simultaneously the free energy of the buffers increases by $\kb T\ln2$ (see appendix \ref{sec:appszilardextract} for more details on this calculation).

At the end of the cycle, both the memory molecule and the input have been returned to unbiased and statistically uncorrelated states. Chemical free energy has been transferred from the buffers 1 and 2, to buffers 3 and 4. The net chemical work extracted is then zero since the $\kb T\ln2$ cost of measurement balances the work extracted. This is, of course, expected---extracting work from the initially equilibrated input should be impossible. However, this basic design will underpin that of devices intended to exploit structured environments, and recover net positive work. 

We note, in passing, \teo{three} instructive features of our explicitly-described biochemical Szilard engine. \teo{Firstly, the measurement and feedback reactions can be implemented sequentially by coupling to  buffers of first one, then another, type of fuel molecule. This ability to switch from having the input set the memory, to having the memory modulate the evolution of the input, is the key feature of our setup that allows us to represent the full cycle.}
Secondly, there is no need for an `erase' step to reset the memory to a specific state \cite{bennett1982thermodynamics}. Whilst it would be possible to include such a reset, it is not necessary, either for efficient operation or to preserve the second law of thermodynamics. The second law is preserved simply by the `ominous' nature of the non-equilibrium correlations originally identified by Szilard. Thirdly, the measurement is simply the act of setting the engine into the correct state to exploit the input (setting the memory to $\m[0]$ or $\m[1]$). There is no need for any other system, intelligent or otherwise, to record or be aware of the outcome of the measurement. In the context of the typical one-particle-gas description of Szilard's engine \cite{szilard1929entropieverminderung}, the measurement is simply the correlation of the pulley and particle positions. Any additional recording of the particle position (for example in the  brain of an intelligent being) corresponds to a useless extra correlation or measurement, with associated costs that must be carefully recovered at a later time to reach 100\% efficiency. 

\subsection{Exploiting a series of correlated bits}
Although the Szilard engine cannot extract useful work from its equilibrium input, it forms the basis of a device for exploiting a series of identical biochemical bits labelled with the index $i$, whose correlated states, described by the random variables $\{X_i\}$, are generated by a stationary stochastic process. The random variable $X_i$ has the possible outcomes of $\x$ or $\xs$. We consider the series to be infinite in both directions. As with the Szilard engine in section \ref{sec:szilard}, both states of the input bits are assumed to be equally intrinsically stable, and separate bits do not interact (they are in different boxes in the language of figure \ref{fig:setup}). The equilibrium distribution of the inputs is, then, for each molecule to be independently distributed uniformly between its two states.

Free energy is stored in the input array if either an initial bias towards $\x$ or $\xs$ is present, and/or correlations exist between $X_i$ and $X_j$ for $i \neq j$. Since designing a system to exploit an intrinsic bias is simple, and requires no measurement or inference (see section \ref{sec:extract}), we focus exclusively on the case in which the marginalised probability of each bit occupying either state is $1/2$.  

\subsubsection{Bounds on work extraction}
\label{sec:bounds}
The free energy per bit stored in such an array, and hence the available work per bit, is determined by the difference between the equilibrium Shannon entropy per bit of $\ln 2$ and the entropy rate $h$ \cite{boyd2016identifying}
\begin{equation}
	W_\text{available}=\kb T\left(\ln2-h \right),
    \label{eq:wavailable}
\end{equation}
where 
\begin{equation}
    h = \lim_{n\rightarrow\infty}\frac{H(X_1,X_2,\dots,X_n)}{n}.
    \label{eq:entropyrate}
\end{equation}

An array of $N$ bits has a state space of size $2^N$. For an array with arbitrary correlations, an operation must be `globally integrated' across all $N$ bits to fully extract $W_\text{available}$ \cite{boyd2018thermodynamics}.
Even if a system were able to achieve this integration by coupling to all bits in an array simultaneously, extracting the full available work would be highly non-trivial. In practice, the protocol would need to be tuned to the expected initial occupancy of each of the $2^N$ states to avoid losses.

The opposite limit to a device that is able to interact with the entire input at once is a device that interacts with each bit separately and in an independent manner.
However, such a device can only extract the free energy stored in the state $X_i$, $\mathcal{F}_{\text x}(X_i)$, having marginalised over all other $X_{j \neq i}$ . In our setting, $\mathcal{F}_{\text x}(X_i) = \mathcal{F}_{\text x}^{\rm eq}$ and thus no work can be extracted. The correlations are wasted and a `modularity cost' is incurred due to the fact that before the work extraction there is mutual information between $X_i$ and later input states, but after the work extraction that mutual information is zero \cite{boyd2018thermodynamics}.

Let us consider a simple extension to the independent-bit device that is interpretable and offers the potential of extracting at least some of the stored work whilst retaining limited complexity. We still manipulate input bits individually, but allow for a memory that maintains its state when the device moves to the next subsystem. This memory permits some of the free energy stored in correlations between successive inputs to be exploited. We now derive a bound on work extraction by this method. 

Consider two adjacent input bits labelled $i$ and $i+1$, and the memory system. The initial state of the $i$th bit is the random variable $X_i$. $X_i$ can take two values: $\x$ or $\xs$. During the interaction of the memory system with the $i$th bit, ${X}_i$ is both measured and recorded in the memory as the state $M_i$, and work is extracted from the $i$th bit as it relaxes to a state $X_i^\text{final}$. We are now concerned with the work that can subsequently be extracted from the $i+1$th bit following the same procedure, given the correlations between $M_i$ and $X_{i+1}$ induced by the measurement.

Let $\mathcal{F}_{\text{joint}}({X}_{i+1},M_i)$ be the free energy of the joint system consisting of the $i+1$th bit and the memory system when in states $X_{i+1}$ and $M_i$ respectively. Before and after the coupling of the $i+1$th bit and the memory system, there is no direct interaction between the two subsystems, and hence the free energy can be written as the sum of individual contributions calculated using marginalised probabilities and an informational term arising from the correlation between $X_{i+1}$ and $M_i$ as in equation \ref{eq:informationisfreeenergy} \cite{parrondo2015thermodynamics}. Prior to measurement, we have 
\begin{align}
	\mathcal{F}_\text{joint}({X}_{i+1},M_i)=\mathcal{F}_{\text{X}}(X_{i+1})+\mathcal{F}_{\text{M}}(M_i) \nonumber \\
    +k_\mathrm{B}T \mathcal{I}(X_{i+1};M_i).
\end{align}

After the interaction window, we have
\begin{multline}
	\mathcal{F}_\text{joint}({X}_{i+1}^\text{final},M_i)=\mathcal{F}_{\text{X}}(X_{i+1}^\text{final})+\mathcal{F}_{\text{M}}(M_{i+1})\\
	+k_\mathrm{B}T \mathcal{I}(X_{i+1}^\text{final};M_{i+1}).
\end{multline}
The work extracted by any process operating between these start and end points is bounded by
\begin{align}
	W \leq \mathcal{F}_{\text{X}}(X_{i+1})-\mathcal{F}_{\text{X}}(X_{i+1}^\text{final})+\mathcal{F}_{\text{M}}(M_i)-\mathcal{F}_{\text{M}}(M_{i+1})\nonumber\\
    +\kb T\Big(\mathcal{I}(X_{i+1};M_i)-\mathcal{I}(X_{i+1}^\text{final};M_{i+1})\Big).
\end{align}
If the process that produces the inputs is stationary and the measurement protocol is the same each time, then
$\mathcal{F}_{\text M}(M_i)=\mathcal{F}_{\text M}(M_{i+1})$. Moreover, by design, $\mathcal{F}_{\text{X}}(X_{i+1})= \mathcal{F}_{\text{X}}^{\rm eq}$ is minimal since 
the marginalised distribution of ${X}_{i+1}$ is the equilibrium one. Invoking the positivity of the mutual information \cite{Elements_of_Information_Theory}, we see that the available work is maximal when ${X}^\text{final}_{i+1}$ also follows an equilibrium distribution, and the extraction process fully decorrelates the input from the memory ($\mathcal{I}(X_{i+1}^\text{final};M_{i+1})=0$). Thus the work extracted per input bit is bounded by
\begin{equation}
	W \leq \kb T \mathcal{I}(X_{i+1};M_i).
\label{eq:W_limit}
\end{equation}
A system that does not make use of a memory, such as the setup for directly exploiting biased inputs discussed in section \ref{sec:extract}, would therefore extract no work. 

The value of the mutual information in equation \ref{eq:W_limit} depends on the details of the measurement process. The state of the memory system, $M_i$, only depends on the state of the next input, $X_{i+1}$, through the previous input state, $X_i$, so by the data processing inequality the maximum work that can be extracted is 
\begin{align}
    W_{\text{single}}^{\text{max}} &= \kb T \mathcal{I}(X_{i+1};X_i)\nonumber\\
    &= \kb T(\ln2-H(X_{i+1}|X_i)).
    \label{eq:extractwithmemory}
\end{align}
This work is, of course, not greater than the available work in the input. The input is stationary so it is possible to write the entropy rate as \cite{Elements_of_Information_Theory}
\begin{equation}
    h=\lim_{n\rightarrow\infty}H(X_n|X_1,X_2,\dots,X_{n-1}),
\end{equation}
and if we use the fact that the conditional entropy is not increased when conditioning on additional variables then
\begin{align}
    W_{\text{single}}^{\text{max}} &= \kb T(\ln2-H(X_{i+1}|X_i))\nonumber\\
    &\leq \kb T(\ln2-h)=W_\text{available}.
\end{align}
These results are a special case of the `modularity cost' outlined in \cite{boyd2018thermodynamics}.

Single bit memory devices are therefore constrained by the amount of information they carry forward to the next bit in the chain. Note that carrying this information forward is not sufficient---it must also be used effectively during the interaction window. One might assume that there is an inherent trade-off between {\em updating} the memory to be the best possible predictor of the next bit, and {\em using} the memory to make the extraction of work from the current bit as efficient as possible. We will now explore this potential trade-off, and these bounds on work extraction more generally, in the context of two distinct devices in two different types of environment.

\subsubsection{Exploiting a Markovian input}
\label{sec:markkovianinput}
We first consider the case in which the binary input is Markovian. That is, the probability distribution of the state of each input molecule only depends on the state of the previous molecule. Since we consider processes which have no bias to either state 0 or 1, this process is a one parameter model given by the probability of transitioning state from one input to the next.
The entropy of a series of $n$ Markovian random variables is
\begin{align}
	H(X_1,X_2,\dots,X_n)&=\sum_{i=1}^nH(X_i|X_1,X_2,\dots,X_{i-1})\nonumber\\
    &=H(X_1)
    +\sum_{i=2}^nH(X_i|X_{i-1}),
\end{align}
in which we have first used the chain rule for conditional entropies \cite{Elements_of_Information_Theory} and second the Markov property. Therefore, from Equation~\ref{eq:wavailable}, the available work if the Markov chain is stationary is
\begin{equation}
	W^{\rm Markov}_\text{available}=\kb T(\ln2-H(X_{i+1}|X_i)).
    \label{eq:wavailable2}
\end{equation}

Comparing Equations~\ref{eq:extractwithmemory} and \ref{eq:wavailable2}, we see that the maximum work for a single bit memory is equal to the full available work in a Markovian environment: $W^{\text{max}}_\text{single}=W^{\rm Markov}_\text{available}$. We now outline a device that extracts all of this work, both achieving the required measurement accuracy $H(M_i,X_i)=0$ and using this measurement to extract all of $W^{\text{max}}_\text{single}$ for each bit. 

We first note that any update of the memory from $M_{i-1}$ to $M_i$ must occur before the $i$th bit is allowed to evolve. Thermodynamically efficient manipulation of the $i$th bit requires that any protocol is quasistatic, with the $\x \rightleftharpoons \xs$ reactions reaching equilibrium with respect to the control faster than the control is updated. Thus, as soon as  $\x \rightleftharpoons \xs$ transitions are allowed by the control, all memory of the previous state is necessarily forgotten, and subsequent updates of the memory  using the initial value of $X_i$ are impossible. 

At first glance it might then seem impossible to extract all the work stored in this setting. We must apparently pay to update the memory from $M_{i-1}$ to $M_i$  using input $X_i$ to carry information forward, before we are able to use the memory to exploit $X_i$. The recent result of \cite{owen2017number} does not preclude the possibility of extracting all the stored work, but leaves open the possibility that a single additional `hidden' state might be required to circumvent this apparent problem (see appendix \ref{app:embed}). 

In fact, no additional states are required. The solution is to use the information carried forward from the measurement of the previous input, $\mathcal{I}(X_{i};M_{i-1})$, to make a \teo{low work-cost} but faithful measurement of $X_{i}$, ($M_i$) and then to use that measurement to extract $\kb T \ln 2$ of  work from the relaxation of the $i$th bit exactly as in Szilard's engine, see Section \ref{sec:szilard}. \teo{Here, the low work cost is measured relative to the $\kb T \ln 2$ cost of a na\"{i}ve measurement performed without information carried forward from the previous bit.}

\begin{figure}[h]
	\includegraphics[width=\linewidth]{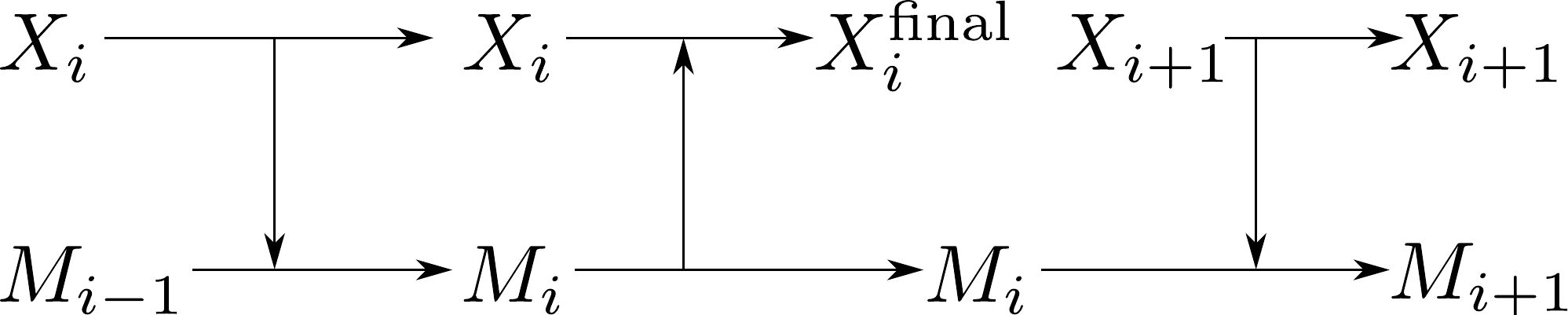}
    \caption{\textit{Schematic diagram of an efficient measurement and feedback process.} The machine receives an input molecule characterised by the random variable $X_i$ and the memory molecule is initially characterised by the random variable $M_{i-1}$. Then, the memory variable is set to $M_i$, which is a measurement of $X_i$. The state $M_{i-1}$ does not affect the state $M_i$ but the fact that $M_{i-1}$ and $X_i$ are correlated means that the measurement can be taken cheaply. Next the correlation between $M_i$ and $X_i$ is used to extract work when changing the state of the $i$th input molecule from $X_i$ to $X_i^\text{final}$, which is in an equilibrium distribution. Then, that input molecule is returned to its box and the next input molecule, with state $X_{i+1}$, is moved to the reaction volume and the process can repeat.}
    \label{fig:markovdiagram}
\end{figure}

An overview of this process is shown in figure \ref{fig:markovdiagram}.
First the new input is copied to the memory. This copy is done using the same chemical reactions as in the measurement step of the biochemical Szilard engine in section \ref{sec:szilard}, repeated here for convenience
\begin{align}
	\m[0]+\x+\fs[1]&\rightleftharpoons\m[1]+\x+\f[1],\nonumber\\
	\m[0]+\xs+\fs[2]&\rightleftharpoons\m[1]+\xs+\f[2].
    \label{eq:markovmeasure}
\end{align}
The only difference from section \ref{sec:szilard} is that now the initial state of input and memory is different. It is still the case that the input molecule and memory molecules are each, when treated in isolation, equally likely to be in both of their states. Now, however, the states of the two molecules are correlated since the memory molecule has been set using the state of the previous input molecule, $X_{i-1}$. A different measurement protocol is therefore needed to make an optimal (reversible) measurement. Instead of starting from a chemical potential difference $\g[1] = 0$ for the fuels, we must start with either $\f[1]$ and $\fs[1]$  in excess so that the equilibrium distribution dictated by this buffer matches the actual biased probability distribution of the memory molecule given that the input molecule is $\x$. Similarly, either $\f[2]$ and $\fs[2]$ must be in excess so that the equilibrium distribution dictated by this buffer matches the biased  probability distribution of the memory molecule given that the input molecule is $\xs$.

The ideal protocol therefore proceeds as follows. Initially, as in the biochemical Szilard engine in section \ref{sec:szilard}, $[\f[1]]$, $[\fs[1]]$, $[\f[2]]$ and $[\fs[2]]$ are all set to zero so the reactions in equation \ref{eq:markovmeasure} cannot occur. Then, the concentrations are simultaneously increased at a fixed ratio of $[\f[1]]/[\fs[1]]$ and $[\f[2]]/[\fs[2]]$ that maintains constant free-energy changes for the reactions of equation \ref{eq:markovmeasure}, $\g[1]=\mu_{\f[1]}-\mu_{\fs[1]}$ and $\g[2]=\mu_{\f[2]}-\mu_{\fs[2]}$, such that
\begin{align}
	\frac{\e{-\beta\g[1]}}{1+\e{-\beta\g[1]}}&=P(M_{i-1}=\m[1]|X_i=\x) \nonumber\\
&=P(X_{i-1}=\xs|X_i=\x),\nonumber\\
	\frac{\e{-\beta\g[2]}}{1+\e{-\beta\g[2]}}&=P(M_{i-1}=\m[1]|X_i=\xs) \nonumber\\
&=P(X_{i-1}=\xs|X_i=\xs).
\end{align}
The reactions catalysed by whichever of $\x$ or $\xs$ is present now occur at an appreciable rate, but forwards reactions exactly balance reverse reactions so there is no overall change in the probability of observing of $\m[1]$ and $\m[2]$. If there is no overall bias towards $\x$ or $\xs$ then $\g[1] = -\g[2]=\g[\text{offset}]$ by symmetry. We have used the term `$\g[\text{offset}]$' because the chemical potential difference has been `offset' from zero, which is what it would be if the successive input molecules were uncorrelated.

The rest of the protocol is the same as for the measurement step of the biochemical Szilard engine in section \ref{sec:szilard}. Next, $[\f[1]]$ and $[\fs[2]]$ are increased while $[\fs[1]]$ and $[\f[2]]$ are kept constant until $[\f[1]]\gg[\fs[1]]$ and $[\fs[2]]\gg[\f[2]]$. Then, $[\f[1]]$, $[\fs[1]]$, $[\f[2]]$ and $[\fs[2]]$ are decreased while maintaining $[\f[1]]\gg[\fs[1]]$ and $[\fs[2]]\gg[\f[2]]$ until $[\fs[1]]=[\f[2]]=0$. Finally $[\f[1]]$ and $[\fs[2]]$ are decreased to zero. Now the reactions in equation \ref{eq:markovmeasure}, again, cannot occur so the memory molecule is fixed to be $\m[0]$ if the input is $\x$ and $\m[1]$ if the input is $\xs$.

The work done by the chemical fuel baths to make this measurement is once more calculated as in section \ref{sec:extract} but with different limits on the integral due to the different $\g[\text{offset}]$. As shown in appendix \ref{sec:appmarkovmeasure}, the work done is exactly $\kb TH(M_{i-1}|X_i)=\kb TH(X_{i-1}|X_i)$, as expected from the change in entropy of the input molecule and memory molecule joint system.

Now that the state of the memory molecule has been updated so that $M_i$ perfectly reflects  $X_i$, $\kb T\ln2$ work is extracted in exactly the same way as in the biochemical Szilard engine. Thus  the net work extracted per input molecule is
\begin{equation}
	W^\text{Markov}=\kb T(\ln2-H(X_{i+1}|X_i)) = W^\text{max}_\text{single} = W^\text{Markov}_{\rm available},
	\label{eq:markovmachinenet}
\end{equation}
which is all the available work in a stationary Markovian input as in equation \ref{eq:wavailable2}. This machine has 100\% efficiency and there is no irreversible dissipation.

It is therefore possible that a machine with a two-state memory that is well-calibrated for this Markovian environment---with the correct initial $\g[\text{offset}]$ in chemical potentials to reflect the nearest-neighbour correlations in the Markov chain---can extract all of the work available. Such a machine faces no trade-offs between exploiting and measuring $X_i$; the exact measurement of $X_i$ both carries the maximal information forward, and enables its full exploitation.
How a machine might obtain the optimal offset parameter, either via design or some form of evolution (to effectively infer the one parameter specifying the Markov process), is beyond the scope of this paper. \teo{We emphasise that calibration of $\Delta G_{\rm offset}$ to the environment is not equivalent to being tuned to the specific fluctuations of one realisation of the environment, but rather to the overall statistical properties of the fluctuations. A poorly-chosen parameter would result in the  `mismatch costs' identified by Kolchinsky and Wolpert \cite{kolchinsky2017dependence}.} 

\subsubsection{Exploiting a non-Markovian input}
\label{sec:nonmarkkovianinput}
\begin{figure}[!]
	\includegraphics[width=\linewidth]{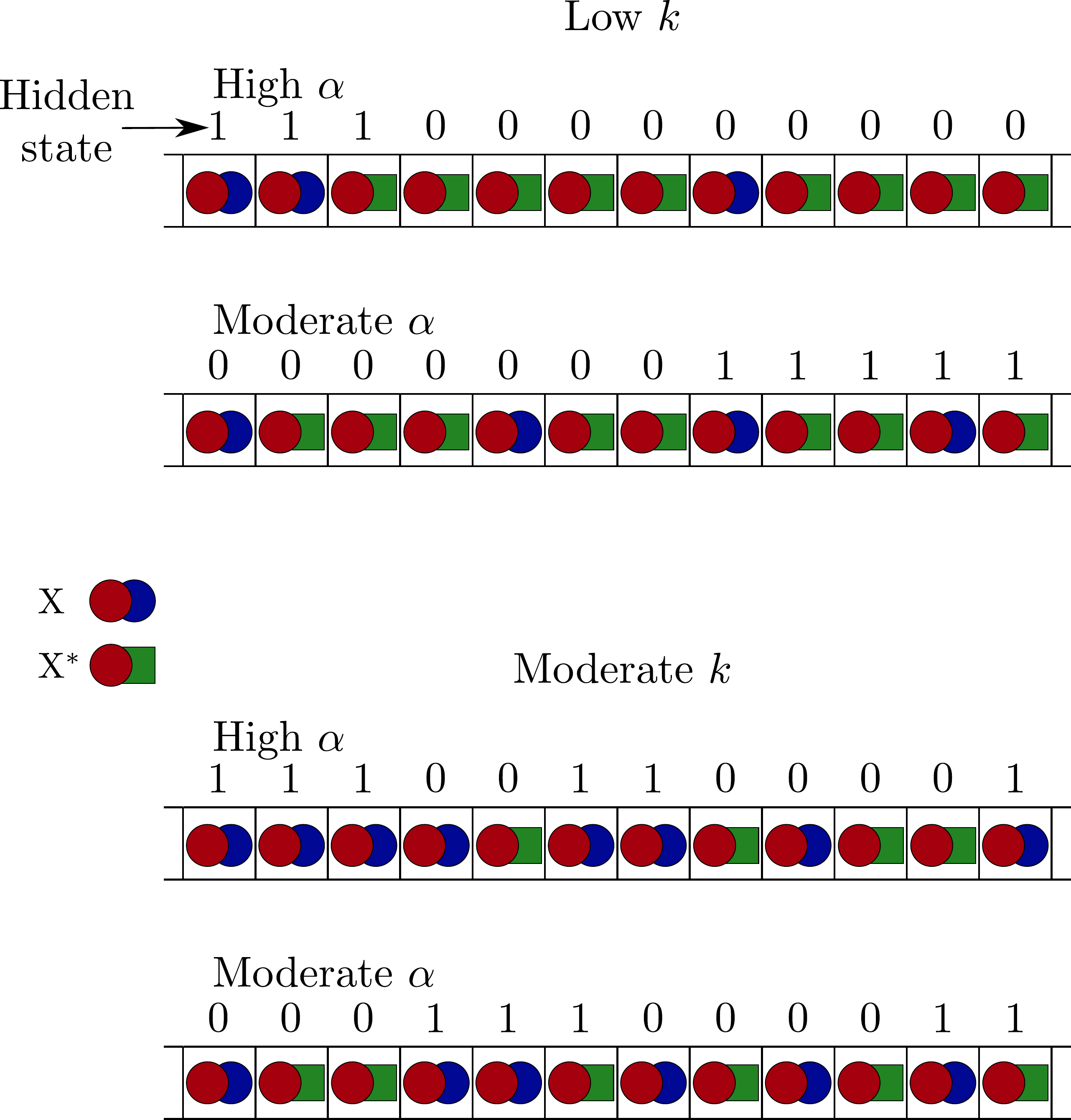}
    \caption{\textit{Samples of inputs produced by the hidden Markov model for various values of the parameters.} When $k$ is small, the hidden state has a low probability of changing at each step so there are long strings \teo{in which the system is biased towards either $\x$ or $\xs$. When $k$ is higher then the hidden state changes more frequently, and the bias is more short-lived}. When $\alpha$ is high then the state of the input molecule is \teo{strongly biased} to match the hidden state, whereas when $\alpha$ is smaller, there is a weaker correlation between the hidden state and the state of the input molecule.}
    \label{fig:hmm}
\end{figure}

In a Markovian environment, if a machine measures the state of an input molecule it knows everything it could about the distribution of the next input. A more complex environment might have correlations that are not fully-described by those of adjacent inputs. In particular, we might imagine an environment with a hidden state $S_i$ that influences the probability of $X_i$; as the hidden state changes, the device moves between regions in which the apparent environmental bias is different. The machine's challenge then becomes a more obvious inference task: to infer the overall state of the environment, and to accordingly exploit the inputs.

\begin{figure}[!]
	\includegraphics[width=\linewidth]{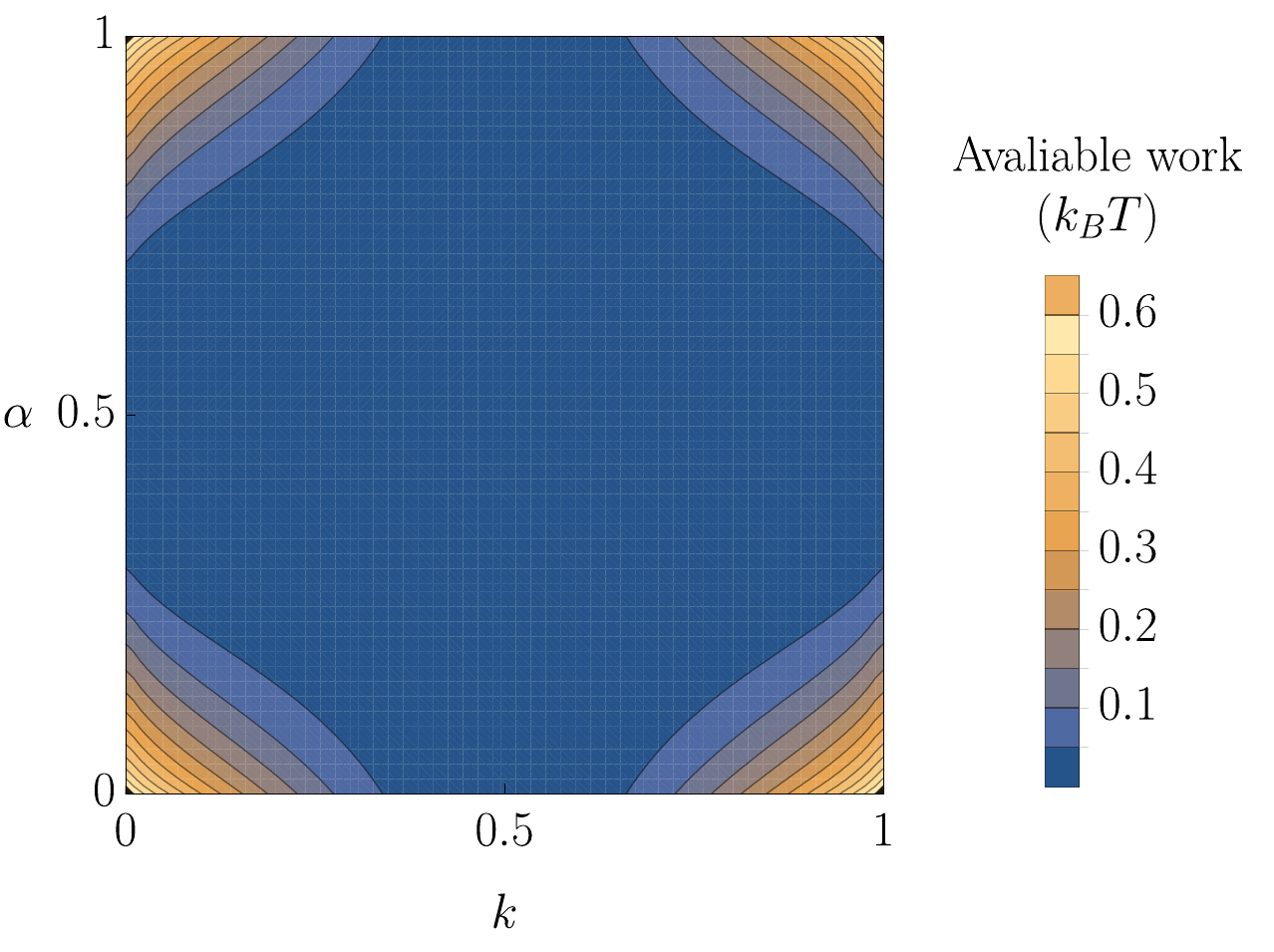}
    \caption{\textit{Available work per molecule for the hidden Markov model input.} The available work is calculated \teo{as a function of bias $\alpha$ and persistence of the hidden state $k$} by numerically evaluating equations \ref{eq:wavailable} and \ref{eq:entropyrate} for $n=20$ and is given in units of $\kb T$.}
    \label{fig:hmm_available}
\end{figure}

Specifically, we will consider a hidden state $S_i$ with $s_i \in \{0,1\}$. When moving from one input molecule to the next the hidden state has a probability $k$ of changing. Conditioned on the hidden state, each input molecule is an independent Bernoulli random variable. The probability of an $\xs$ molecule is $\alpha$ if $S_i=0$ and $1-\alpha$ if $S_i=1$. Some example sequences produced by this process are shown in figure \ref{fig:hmm}. Due to the overall symmetry of the process, $X_i =\x$ and $X_i = \xs$ are both equally likely having marginalised over all inputs $j \neq i$. Thus, as in section \ref{sec:markkovianinput}, no free energy is stored in the state of single molecules---only in the correlations between different molecules. The available work that can be extracted per input molecule is plotted against the parameters $k$ and $\alpha$ in figure \ref{fig:hmm_available}. Hidden states that either reliably persist ($k \rightarrow 0$) or switch ($k \rightarrow 1$), and which provide a predictable output ($\alpha \rightarrow 0,1$) lead to the most free energy stored in the environment.



Given the history of the inputs $\{X_{j<i}\}$, the optimal statistical prediction of the next input $X_i$ can be made via the forward algorithm \cite{stratonovich1960conditional}. A  machine capable of both iterating the forward algorithm at each step, and using the previous value to optimally exploit the current input, would be able to extract the full $W_{\text{available}}$. However, implementing the forward algorithm is impossible for our machine with a single bit of memory that can make only a binary `decision' during its feedback. For a hidden Markov process, the conditional probability distribution of the next input molecule given the entire history of the input is different for all possible states of the history. Equivalently, the process $\{ X_i\}$ cannot be described  by a finite state $\epsilon$-machine \cite{shalizi2001computational}, and thus the forward algorithm requires a memory that is a real number, and the exploitation step would need to have a continuous dependence on this real number. 

It might be tempting to think that a simpler alternative to the forward algorithm, in which the two-bit memory variable $M_i$ is set based on both the current input variable $X_i$ and its previous value $M_{i-1}$, would give better predictions by allowing the machine to take in more historical information at each step. Such an approach would represent a trade-off, with a maximal information carried forward $\mathcal{I}(X_{i+1};M_i)$ being obtained only at the expense of an increased uncertainty $H(X_i,M_i)$ in the state of the current input after the measurement. Whether or not the reduced measurement cost could compensate for the reduction in work obtained during the extraction step is moot, however, since such a strategy is impossible, at least in the quasistatic setting. One cannot update the memory from $M_{i-1}$ to $M_i$ quasistatically, in a way such that $\mathcal{I}(M_i;M_{i-1}) \neq 0$, without access to additional hidden memory states \cite{wolpert2017minimal,owen2017number}. All information on initial conditions is necessarily lost immediately when a degree of freedom evolves under a quasistatic process. Thus in the quasistatic setting at least, our single bit memory cannot trade off the accuracy of measurement of the current input and information carried forward.

\subsubsection{Markov machines in non-Markovian environments}
\label{sec:markovmachinehmmwork}
With the above limitations in mind, we first ask how well the {\em Markov machines} considered in Section \ref{sec:markkovianinput}, that are limited  to interact with one bit at a time, and carry only one bit of memory forward, function in the non-Markovian environment specified.  For a perfect measurement of each bit, such that $H[X_i|M_i]=0$, the expected work extracted per molecule for a quasistatically-operated device still follows from Equation~\ref{eq:extractwithmemory}
\begin{equation}
	W^{\rm max}_\text{single}=\kb T\Big(\ln2-H(X_{i+1}|X_i)\Big),
\end{equation}
but now $W^{\rm max}_\text{single} < W_\text{available}$ since there is additional information in long range correlations that is not taken into account by the information between nearest neighbours. Therefore, the machine has efficiency $\eta =W^{\rm max}_\text{single}/W_\text{available}< 1$ and irreversibly generates entropy.

\begin{figure*}
	\centering
	\includegraphics[scale=0.8]{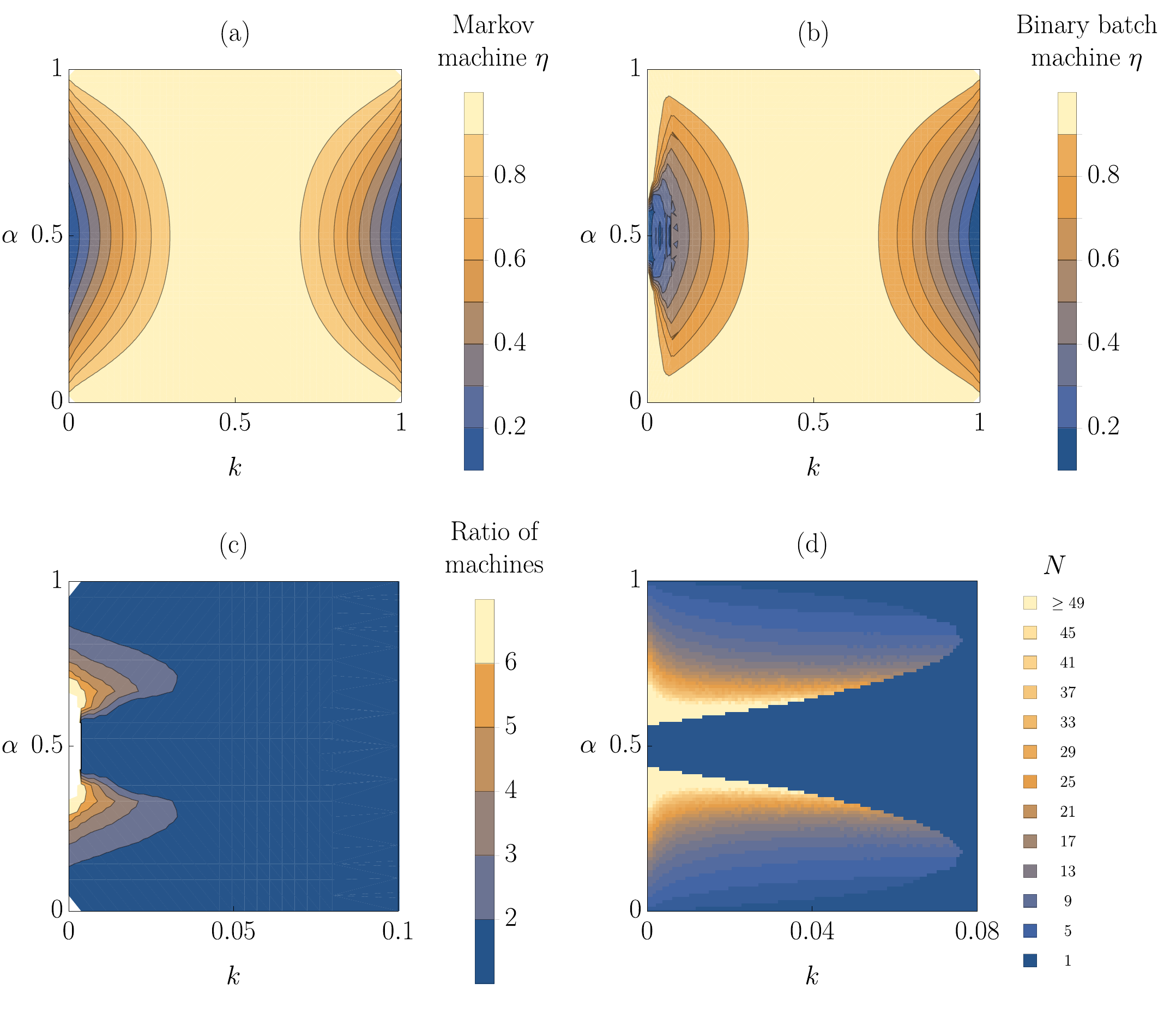}
    \caption{\textit{More complex inference strategies can be beneficial in more complex environments.} (a) The efficiency $\eta$ of the Markov machine when acting on an input produced by a hidden Markov model. The efficiency is low when \teo{the hidden state switching probability} $k$ is close to 0 or 1 but  $\alpha$, \teo{the bias implied by a given hidden state,} is not close to 0 or 1 because there are long range correlations that are masked by short-term fluctuations. The available work is calculated as in figure \ref{fig:hmm_available} and we have an analytic expression for work extracted by the Markov machine from equation \ref{eq:markovmachinenet}. (b) The efficiency of the binary batch machine. The efficiency is the same as the Markov machine except for the region close to $k=0$, where it shows enhanced work extraction. The work extracted by the binary batch machine is calculated by simulating the hidden Markov model for 10,000,000 steps and calculating the work extracted per molecule for the binary batch machine with $N$ from 1 to 25 and selecting the greatest work extracted. (c) The ratio between the work extracted by the binary batch machine and the Markov machine per molecule, clarifying the enhanced performance near to $k=0$. (d) The batch size that gives the greatest expected work as a function of the input parameters. For $k \gtrsim 0.077$ it is always optimal to interact with one molecule at a time; larger batches are favourable for $k \rightarrow 0$ and values of $\alpha$ away from 0, 0.5 and 1. It is always the case that the optimum batch size is odd.}
    \label{fig:efficiencyplots}
\end{figure*}

This efficiency, $\eta$, of Markov machines acting on a hidden Markov model input is plotted in figure \ref{fig:efficiencyplots}(a). In making these plots, we first identify the optimal Markov machine offset parameter at each $\alpha$ and $k$, and then calculate the efficiency of that device---once again assuming that the machine's parameters are optimised to the statistical properties of its environment (perhaps through evolution). It is notable that the Markov machines perform reasonably well in these environments, except when $k \rightarrow 0$ or 1, and $\alpha  \not\approx 0$ or 1. In these environments, the hidden state behaves predictably and so correlations  are long-ranged, but $X_i$ fluctuates considerably within the hidden state, effectively fooling the Markov machine that is only able to predict $X_i$ based on $X_{i-1}$. 

The behaviour of the Markov machine can be related to that of the Kalman filter \cite{kalman1960new}, an algorithm for making real-time predictions of the state of a noisy dynamical system with noisy measurements of the system's state. The relative weight put on previous measurements versus the most recent input is a parameter that can be adjusted, and it is well known that high intrinsic noise implies that the current measurement should be weighted strongly, whereas high measurement noise calls for greater emphasis on the previous measurements. The Markov machine is effectively constrained to put all of its emphasis on the most recent measurement; it therefore functions better when the `intrinsic' noise of the hidden state is relatively high ($k \sim 0.5$ and $\alpha \sim 0,1$), and worse when the `measurement' noise of the inputs is relatively large ($k \sim 0,1$, $\alpha \sim 0.5$). 

\subsubsection{Batch averaging machines in non-Markovian environments}
\label{sec:batchmachine}
\begin{figure}[!]
	\includegraphics[width=\linewidth]{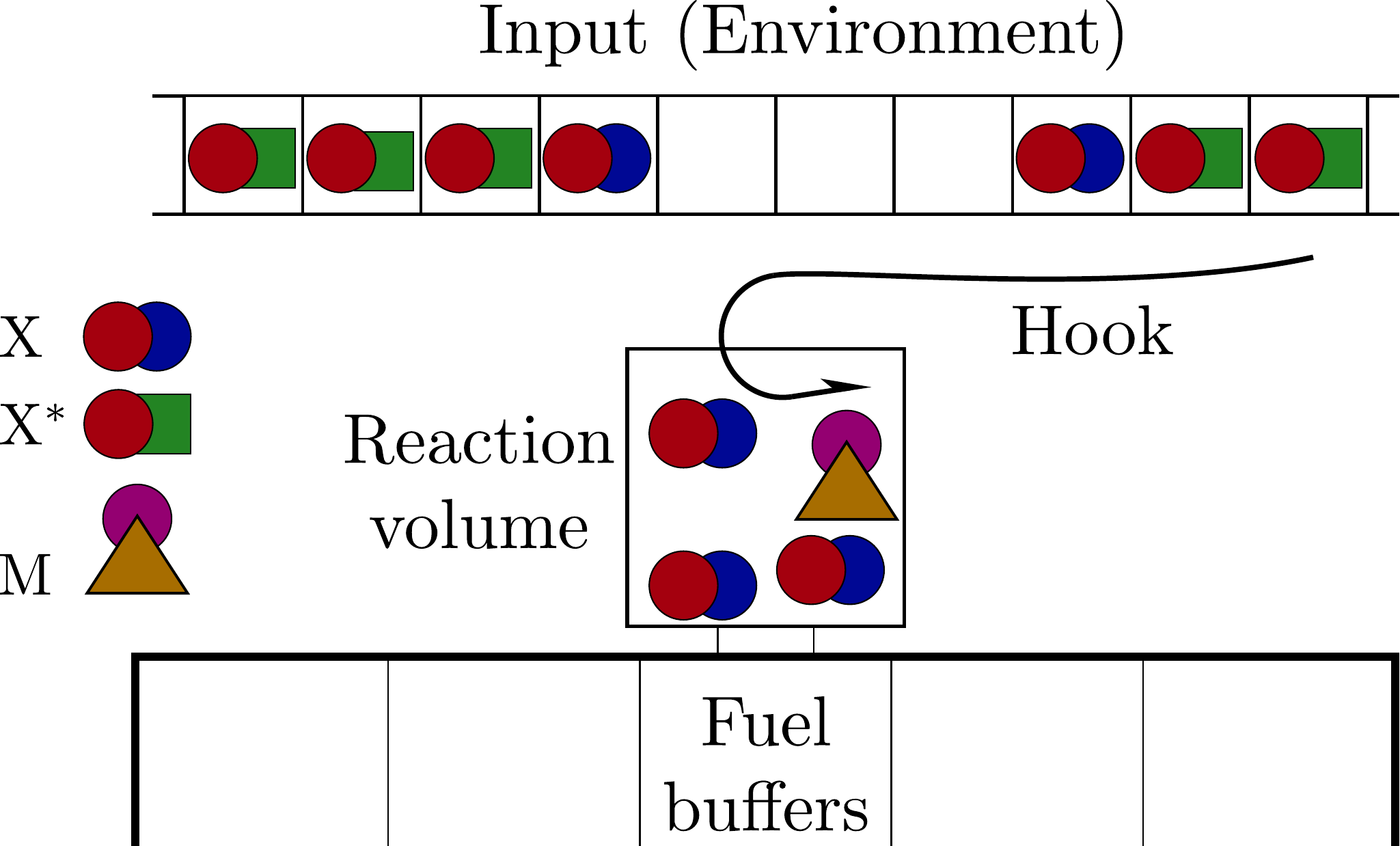}
    \caption{\textit{Schematic of a batch machine for exploitation of non-Markovian environments}. The setup is similar to figure \ref{fig:setup}, so we have omitted the illustration of the control acting on multiple systems simultaneously for clarity. The difference is that now multiple input molecules are moved from their boxes to the reaction volume by the hook. In this figure we have illustrated the case of $N=3$. The input molecules can freely diffuse inside the reaction volume, so their order in the input is lost.}
    \label{fig:batchsetup}
\end{figure}

We now ask whether a more sophisticated strategy, still involving only a single memory molecule and a single binary decision, can overcome this weakness of the Markov machine. If we consider the region where $k\approx0$, then it is likely that a run of multiple input molecules will be produced by the same hidden state. Inspired by our analogy with the Kalman filter, we look for a mechanism of somehow considering multiple input molecules to provide more reliable information about the hidden state, allowing more efficient work extraction. Indeed, in the context of cellular sensing of the concentration of external ligands
 \cite{govern2014optimal, berg1977physics}, it has been observed that averaging approaches can be beneficial when correlation times in the environment are long \cite{Malaguti2018Theory}.


We therefore introduce the {\em batch machine}, illustrated in figure \ref{fig:batchsetup}, which is similar to the Markov machine except that it interacts with (i.e. measures and exploits) a batch of multiple molecules simultaneously, rather than just one. An $N$-batch machine operates by: (a) transferring $N$ inputs to the reaction volume (with no work cost---see appendix \ref{sec:apphook}); (b) performing an operation to set a memory based on these $N$ inputs (for a low work cost because the state of the batch is correlated with the state of the memory, which is set based on the state of the previous batch); (c) exploiting the $N$ inputs simultaneously using the memory; and (d) transferring the $N$ inputs back to their array in a random order. 

We will first consider a `binary' machine that, like the Markov machine, has only two measurement reactions and two work extraction reactions. Let $J_i$ be the random variable representing whether  the number of $\xs$ molecules in batch $i$ is greater than half the batch-size, $N/2$ ($J_i=1$ if true, 0 otherwise). The machine performs measurement of batch $i$ by setting the memory molecule to $M_i = \m[0]$ if $J_i=0$, and to $M_i =\m[1]$ if $J_i=1$; we note that other binary measurement choices are possible, but this simple one serves to illustrate the possibilities of a more complex inference strategy. The machine then exploits the imbalance of inputs in the same way that the Markov machine exploits a measured $X_i = \x$ or $X_i = \xs$, by allowing the inputs to relax to an unbiased distribution whilst transferring free energy to chemical buffers. For $N=1$, the binary batch machine is identical to the Markov machine of section \ref{sec:markkovianinput}; for $N>1$ the initial measurement essentially performs an average over $N$ inputs to set its memory. In the limit $N\rightarrow \infty$, the batch machine interacts with all molecules at once. However, with only two possible measurement states (and hence two possible work extraction strategies), this limit is generally inefficient. 

The measurement can be done with the reactions
\begin{align}
	\m[0]+\frac{N+1}{2}\x+\fs[1]&\rightleftharpoons\m[1]+\frac{N+1}{2}\x+\f[1], \nonumber\\
    \m[0]+\frac{N+1}{2}\xs+\fs[2]&\rightleftharpoons\m[1]+\frac{N+1}{2}\xs+\f[2],
    \label{eq:batchmeasureodd}
\end{align}
when $N$ is odd and
\begin{align}
	\m[0]+\frac{N}{2}\x+\fs[1]&\rightleftharpoons\m[1]+\frac{N}{2}\x+\f[1],\nonumber\\
    \m[0]+\left(\frac{N}{2}+1\right)\xs+\fs[2]&\rightleftharpoons\m[1]+\left(\frac{N}{2}+1\right)\xs+\f[2],
    \label{eq:batchmeasureeven}
\end{align}
when $N$ is even. There is an $N/2$ in one of the reactions in equation \ref{eq:batchmeasureeven} and an $N/2+1$ in the other because for even $N$ there is an arbitrary choice as whether to assign the state where there are $N/x$ molecules of  $\xs$ in the batch to $\m[0]$ or $\m[1]$. We have chosen to assign the state with $N/2$ $\xs$ molecules to $\m[0]$. Clearly, if there are $N$ molecules of $\x$ and $\xs$ in total, then only one of these reactions in equations \ref{eq:batchmeasureodd} and \ref{eq:batchmeasureeven} can occur at once. \teo{As a result, if the reactions are once more driven in opposite directions by fuel imbalances, an excess of $\x$ molecules can be used set the memory to $\m[0]$, and an excess of $\xs$ can be use to set the memory to $\m[1]$.} We immediately see the price for a more complicated strategy---our reactions now require $\sim N/2$ molecules to act as combined catalysts, rather than just a single input molecule (see appendix \ref{sec:appbatchdna} for a DNA strand displacement design for these reactions).

The protocol of changes to the fuel molecule concentrations required for measurement is very similar to that of the Markov machine. Initially, $[\f[1]]$, $[\fs[1]]$, $[\f[2]]$ and $[\fs[2]]$ are all set to zero so the reactions in equations \ref{eq:batchmeasureodd} and \ref{eq:batchmeasureeven} cannot occur. Then, the concentrations are simultaneously increased at a fixed ratio of $[\f[1]]/[\fs[1]]$ and $[\f[2]]/[\fs[2]]$ that maintain overall free energy changes for the reactions, $\g[1]=\mu_{\f[1]}-\mu_{\fs[1]}$ and $\g[2]=\mu_{\f[2]}-\mu_{\fs[2]}$, such that
\begin{align}
	\frac{\e{-\beta\g[1]}}{1+\e{-\beta\g[1]}}&=P(M_{i-1}=\m[1]|J_i=0)\nonumber\\
	&=P(J_{i-1}=1|J_i=0),\nonumber\\
	\frac{\e{-\beta\g[2]}}{1+\e{-\beta\g[2]}}&=P(M_{i-1}=\m[1]|J_i=1),\nonumber\\
	&=P(J_{i-1}=1|J_i=1).\
\end{align}
In exactly the same way as in the biochemical Szilard engine and in the Markov machine, depending on whether there are more $\x$ or $\xs$ molecules in the batch, one of the measurement reactions can now occur at an appreciable rate. The initial offsets $\g[1]$ and $\g[2]$ allow information between batches to be exploited, and are exactly analogous to the constant offsets introduced in section \ref{sec:markkovianinput}. Like in the Markov machine, if $N$ is odd then $\g[1]=-\g[2]=\g[\text{offset}]$ by symmetry. If $N$ is even then $\g[1]\neq-\g[2]$ because $P(J_i=0)\neq P(J_i=1)$.

Then, as in the previous machines, $[\f[1]]$ and $[\fs[2]]$ are increased while $[\fs[1]]$ and $[\f[2]]$ are kept constant until $[\f[1]]\gg[\fs[1]]$ and $[\fs[2]]\gg[\f[2]]$. Subsequently, $[\f[1]]$, $[\fs[1]]$, $[\f[2]]$ and $[\fs[2]]$ are decreased while maintaining $[\f[1]]\gg[\fs[1]]$ and $[\fs[2]]\gg[\f[2]]$ until $[\fs[1]]=[\f[2]]=0$. Finally $[\f[1]]$ and $[\fs[2]]$ are decreased to zero. Now the reactions in equations \ref{eq:batchmeasureodd} and \ref{eq:batchmeasureeven}, again, cannot occur and the memory molecule has been set to state $\m[0]$ if the batch contains more $\x$ molecules that $\xs$ or equal number of $\x$ and $\xs$ molecules, and to state $\m[1]$ if the batch contains more $\xs$ molecules than $\x$.

The cost of making the measurement is calculated in exactly the same way as for the Markov machine (see appendix \ref{sec:appbinarybatchmeasure}), and gives
\begin{equation}
    W^\text{measure}=-H(J_{i+1}|J_i).
\end{equation}
The negative sign represents negative work extraction.

Subsequently, work is extracted from the correlated state of the measurement molecule and the batch. The binary batch machine uses the same reactions as the biochemical Szilard engine and the Markov machine to extract work; they are repeated here for convenience:
\begin{align}
	\m[0]+\x+\fs[3]&\rightleftharpoons\m[0]+\xs+\f[3],\nonumber\\
	\m[1]+\x+\fs[4]&\rightleftharpoons\m[1]+\xs+\f[4].
    \label{eq:batchextract}
\end{align}
However, the protocol is modified, which is necessary because the state of the memory molecule does not report perfectly on the state of the inputs: any number of molecules in state $\xs$ greater than $N/2$ correspond to $J_i=1$ and hence $M_i=\m[1]$, for example.  Initially $[\f[3]]=[\fs[3]]=[\f[4]]=[\fs[4]]=0$, along with the fuels used in the measurement process. Then, the concentrations are simultaneously increased at a fixed ratio of $[\f[3]]/[\fs[3]]$ and $[\f[4]]/[\fs[4]]$ that maintain an overall free energy changes for the reactions, $\g[3]=\mu_{\fs[3]}-\mu_{\f[3]}$ and $\g[4]=\mu_{\fs[4]}-\mu_{\f[4]}$, such that
\begin{align}
	\frac{\e{-\beta\g[3]}}{1+\e{-\beta\g[3]}}&=\hat{p}_0,\nonumber\\
	\frac{\e{-\beta\g[4]}}{1+\e{-\beta\g[4]}}&=\hat{p}_1,
	\label{eq:binarybatchpotentials}
\end{align}
where $\hat{p}_0$ is the probability that an input molecule in the batch is in the state $\xs$, conditioned on  $J_i=0$, and $\hat{p}_1$ is the probability that an input molecule in the batch is in the state $\xs$ conditioned on $J_i=1$. It is clear that $[\f[3]]>[\fs[3]]$ { (assuming that the $\f[3]$ and $\fs[3]$ molecules have equal free energy)} because $\hat{p}_0<1/2$ and $[\f[4]]<[\fs[4]]$ because $\hat{p}_1>1/2$. On average, no chemical work is done at this stage. Similar to the measurement, if $N$ is odd then $\g[3]=-\g[4]$ by symmetry. If $N$ is even then $\g[3]\neq-\g[4]$ because $\hat{p}_0\neq1-\hat{p}_1$.

Work can then be extracted from the batch exactly as for $N$ independent molecules with a bias represented by $\hat{p}_0$ and $\hat{p}_1$. Therefore, $[\fs[3]]$ and $[\f[4]]$ are increased quasistatically until the free energy differences, $\g[3]$ and $\g[4]$, are zero. Then $[\f[3]]$, $[\fs[3]]$, $[\f[4]]$ and $[\fs[4]]$ are decreased to zero while maintaining the same $[\f[3]]/[\fs[3]]$ and $[\f[4]]/[\fs[4]]$ ratios. When $[\f[3]]=[\fs[3]]=[\f[4]]=[\fs[4]]=0$ again the reactions in equation \ref{eq:batchextract} cannot occur. So, finally, the batch reaches an unbiased equilibrium, and during this process the free energy of the buffers are increased. The work extracted in this step is simply $N$ times the work extracted from one input molecule with a bias of $\hat{p}_0$ if $J_i=0$  or $\hat{p}_1$ if the number of $J_i=1$. It is therefore
\begin{align}
	W_\text{batch}^\text{extract}=P(J_i=0)\kb TN&(\ln2+\hat{p}_0\ln\hat{p}_0\nonumber\\
    &+(1-\hat{p}_0)\ln(1-\hat{p}_0))\nonumber\\
    P(J_i=1)\kb TN&(\ln2+\hat{p}_1\ln\hat{p}_1\nonumber\\
    &+(1-\hat{p}_1)\ln(1-\hat{p}_1)).
\end{align}
Therefore, the net work extracted by the binary batch machine from one batch is
\begin{multline}
	W_\text{batch}=\kb TN\Big(\ln2\\
	+P(J_i=0)\big(\hat{p}_1\ln \hat{p}_1+(1-\hat{p}_1)\ln(1-\hat{p}_1)\big)\\
    +P(J_i=1)\big(\hat{p}_2\ln\hat{p_2}+(1-\hat{p}_2)\ln(1-\hat{p}_2)\big)\Big)\\
    -\kb TH(J_{i+1}|J_i)].
\end{multline}

As with the Markov machine, we can ask the question of how the optimal batch machine (with $N$ and the free-energy offsets of the fuel baths optimally tuned to the environmental parameters $k$ and $\alpha$), would perform. Note that since the binary batch machine with $N=1$ is a Markov machine, the optimal binary batch machine must perform at least as well as the optimal Markov machine.

 \begin{figure*}[!]
    	\includegraphics[width=\linewidth]{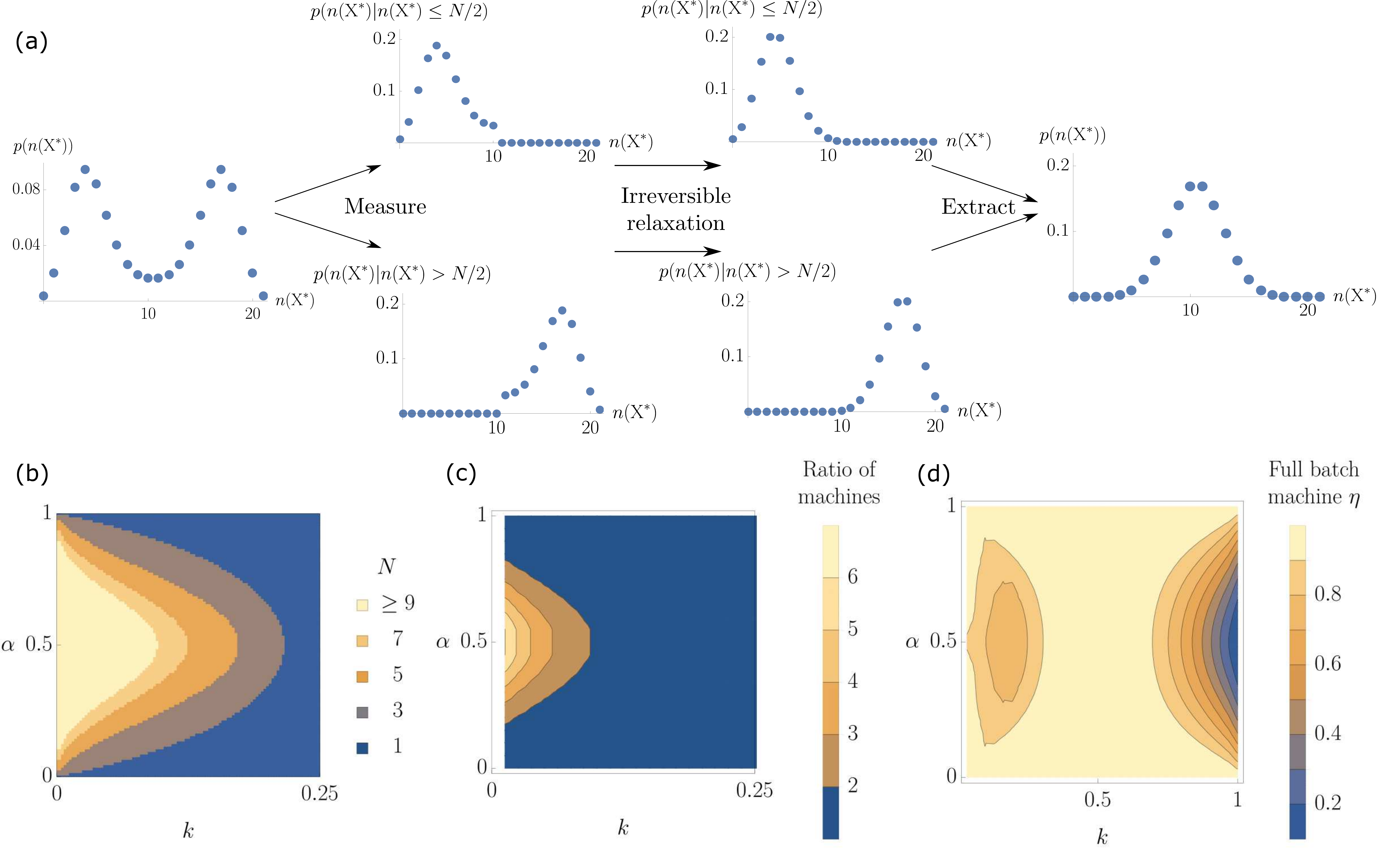}
    \caption{\textit{Inefficiency in the operation of the batch machine arises from uncontrolled relaxations prior to work extraction.} (a) Evolution of the probability distribution of the number of molecules in the $\xs$ state during the operation of the batch machine. { These probabilities can be calculated exactly from the probabilities of each possible output from the hidden Markov model of length 20.} Initially, there is a distribution produced by the hidden Markov model (in this case with $k=0.8$ and $\alpha=0.01$). Then, whether there are more than $N/2$ $\xs$ molecules in the batch is measured, \teo{which produces two possibilities with associated conditional probabilities}. \teo{Next, prior to the work extraction stage of the protocol,} the reaction volume is put in contact with a buffer of fuel molecules with chemical potential differences as defined in equation \ref{eq:binarybatchpotentials}, \teo{which has a ratio of fuel molecule concentrations consistent with the expected number of X and $\xs$ molecules. The batch of input molecules irreversibly relaxes to equilibrium with the buffer, changing to a binomial distribution of $n(\xs)$ from the post-measurement distribution. No work is extracted on average because both distributions have the same average of $n(\xs)$.} Finally, work is extracted by shifting the probability distribution to the equilibrium distribution. A putative `full' batch machine that avoids the loss of this irreversible relaxation can out-perform the batch machine, as shown in (b)-(d). (b) The greatest expected work as a function of the input parameters for the full batch machine does not have an optimum batch size of 1 for $\alpha\approx0.5$. The optimum $N$ has been found by numerically calculating the work extracted for the values of $N$ up to $N=9$ { using the equations in appendix \ref{sec:fullbatchwork}}. (c) The ratio of the work extracted by the full batch machine and the work extracted by the Markov machine. In contrast to the binary batch machine this full batch machine can extract more work than the Markov machine when $k$ is close to 0 and $\alpha$ is close to 0.5. (d) The efficiency $\eta$ of the full batch machine.}
    \label{fig:binarybatchextract}
\end{figure*}

 The efficiency of the optimal binary batch machine is plotted for different values of $\alpha$ and $k$ in figure \ref{fig:efficiencyplots}(b), showing apparently higher efficiency than the optimal Markov machine for some values of $k$ and $\alpha$ as $k \rightarrow 0$. To make this comparison clearer, in figure \ref{fig:efficiencyplots}(c) we have plotted the work extracted by the binary batch machine per molecule divided by the work extracted by the Markov machine per molecule. We see that there are two regions where the binary batch machine extracts more work. Also, in figure \ref{fig:efficiencyplots}(d) we have plotted the optimal batch size for the binary batch machine for the different values of the parameters. For $k>0.08$ the optimum batch size is always $1$ so the Markov machine and the binary batch machine are the same, but for smaller values of $k$ larger batches are frequently favoured. It is always the case that the optimum batch size is odd, since the extraction reactions of the binary batch machine cannot extract work from a batch with equal numbers of $\x$ and $\xs$ molecules.

The binary batch machine delivers, at least in part, on the prospect of improving work extraction from an environment with more complexity. It is unsurprising that a long hidden state life time, $k \rightarrow 0$, is necessary for this advantage to be manifest: the averaging strategy will clearly fare poorly when the hidden state switches rapidly. When $\alpha$ is close to 0 or 1 the state of the input molecule reflects the hidden state with a high probability so the string of input molecules is approximately Markovian, preventing the batch machine from finding a competitive advantage.  The most subtle question, however, is why the binary batch machine does not extract more work than the Markov machine when $\alpha\approx0.5$ and $k \rightarrow 0$. Na\"ively, this regime would seem to be ideal for the batch machine to extract work from weak, but long-lived biases towards either $\x$ or $\xs$. From the perspective of the analogy with Kalman filters, this regime should favour the approach that considers a wide range of inputs, rather than just the most recent. To understand why this intuition fails, we consider where the thermodynamic losses occur during the operation of the binary batch machine.

Several stages of the operation of the optimal binary batch machine are thermodynamically irreversible, resulting in efficiencies $\eta < 1$. They include the point at which the memory is updated without taking into account correlations between \emph{non}-neighbour batches; the point at which the batch of $N$ input molecules are mixed within the reaction volume (figure \ref{fig:batchsetup});  and the point at which the work extraction begins using the measurement molecule. In the first process, a modularity cost is incurred. In the second, mixing causes the positional order within a batch to be lost, reducing our ability to extract work from the sequence of molecules within the batch. All that remains is a non-equilibrium distribution of the number of molecules in each state.  In the third process, this non-equilibrium distribution relaxes further to a binomial distribution for the number of $\xs$ with parameter $\hat{p}_0$ if $J_i=0$ or $\hat{p}_1$ if $J_i=1$, with no work extracted on average during this relaxation, as shown in figure \ref{fig:binarybatchextract}(a).

We can imagine a putative `full batch machine' that could extract all of the work available from the unordered batch, after the initial mixing and measurement. Such a machine would require additional extraction processes to which the memory could couple in each state. The optimal batch size for this full batch machine is plotted in figure \ref{fig:binarybatchextract}(b). We can see that for this machine it is not the case that the optimal batch size is 1 when $\alpha\approx0.5$. The contour plot for this machine is more similar to expectations: as $k \rightarrow 0$ optimal batch size increases for all values $\alpha$. We have also plotted the ratio between the work extracted by the full batch machine and the Markov machine in figure \ref{fig:binarybatchextract}(c), and see that the full batch machine extracts more work than the Markov machine when $k$ is close to 0 and $\alpha\approx0.5$. Thus the reason that the binary batch machine fails to provide an improvement in the vicinity of $\alpha=0.5$ is at least in part because the free energy wasted during the simple binary work extraction mechanism is too large compared to the relatively low amounts of work available (as seen in figure \ref{fig:hmm_available}).

\subsubsection{Robustness}
\label{sec:robustness}
On average, all the machines can extract a positive amount of work from each input molecule or batch of input molecules. However, in a single realisation of the input produced by the stochastic process it is possible for the machines to extract a negative amount of work; i.e. lose free energy, since the prediction of the upcoming state is only probabilistic even in the best case.

Thus, the total work extracted by any machine is a biased random walk. If the machine is unlucky it can receive a fluctuation in the input and get many negative steps with few positive steps. If we are imagining that the machine needs to harvest enough work to power its decision-making, like a biological organism, a fluctuation in its environment where it loses all of its stored free energy would be disastrous. We therefore also consider fluctuations in the work extracted by the machines. If one protocol has a higher expected work extraction but a larger variance it might not be truly better.

\begin{figure}[h]
	\includegraphics[scale=0.8]{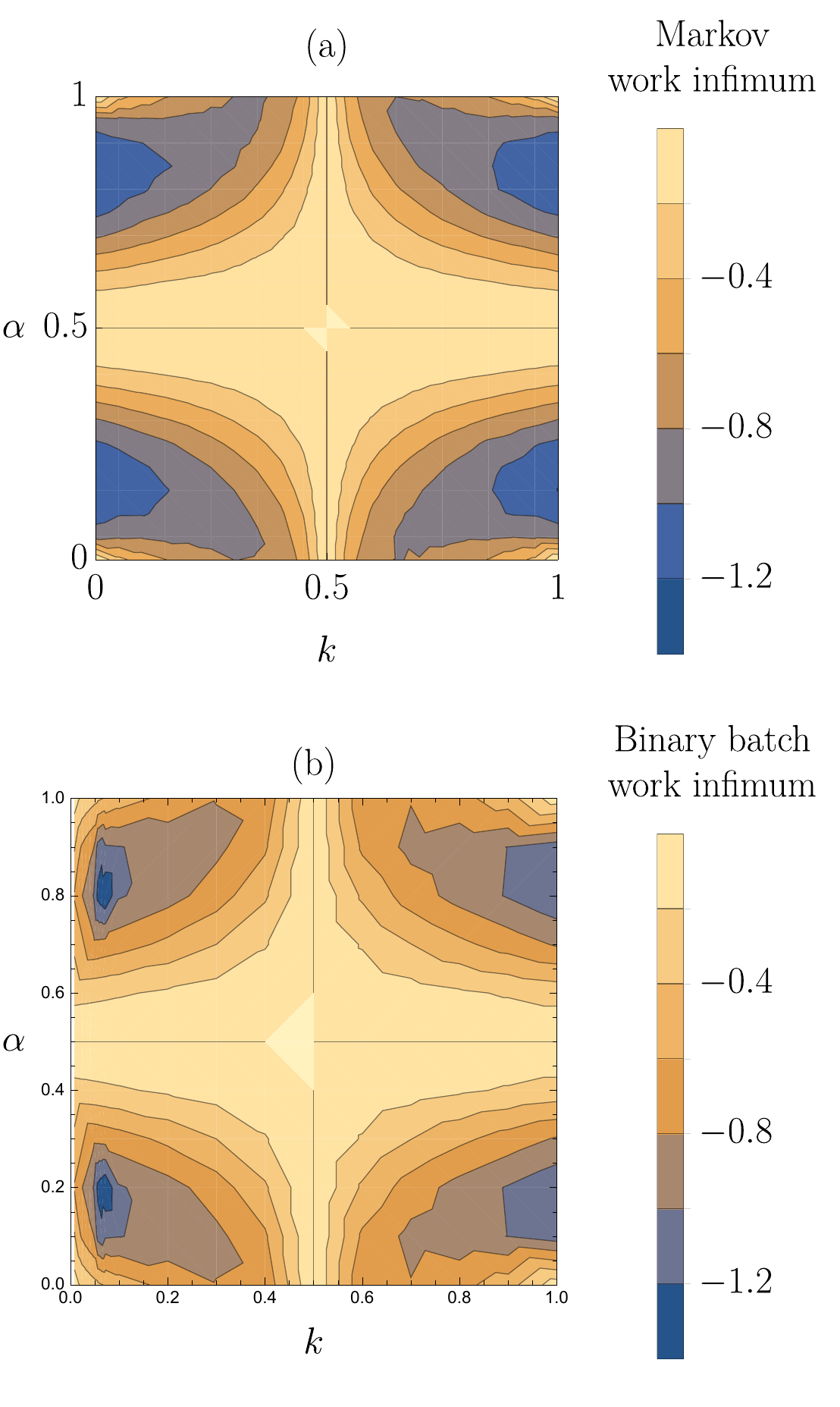}
    \caption{\textit{The binary batch machine is more robust than the Markov machine in the regions in which it extracts more work.} (a) The mean infimum of work of the Markov machine in a run of 100 molecules averaged over 1000 trajectories. (b) The mean infimum of work of the binary batch machine in a run of 100 molecules averaged over 100000 trajectories. We see that in the region where the optimal batch size is greater than 1 (shown in figure \ref{fig:efficiencyplots}(d)) the magnitude of the negative fluctuations is decreased compared to the Markov machine.}
    \label{fig:fluctuationsplot}
\end{figure}

The expected worst-case energy-loss---the infimum of the work extracted---can be thought of as the starting larder-size/fuel-reserves that such a reasoning machine requires. It also gives a minimum timescale that any machine would need to run before it could create a replica that is also robust to environmental fluctuations. This infimum of the total work extracted by the machines in a trajectory averaged over many simulated trajectories is plotted against the parameters of the input process in figure \ref{fig:fluctuationsplot}. When $k\approx0.5$ or $\alpha\approx0.5$, the work that is extracted by the machines is small so the size of the negative fluctuations are also small for both machines. Comparing figures \ref{fig:fluctuationsplot}(a) and (b) shows that the binary batch machine exhibits reduced fluctuations in the regions where $k$ is close to 0 and $\alpha$ is not close to 0, 0.5 or 1, where a batch size greater than 1 is favoured by the average work extracted. This fact is perhaps unsurprising, given that averaging over many inputs is inherently conservative.

\section{Discussion}
We have considered the question of how minimal molecular devices might be designed to exploit the free energy stored in simple non-equilibrium environments. Having outlined a concrete design for a biomolecular Szilard engine, we have shown how such a device can form the basis of machines for exploiting a correlated series of molecular bits, expanding on previous work that has only considered environments with a very particular structure \cite{mcgrath2017biochemical,Stopnitzky2018Physical,chapman2015autonomous}.

Although our devices require externally-applied protocols to operate, all information-processing is performed by degrees of freedom that are explicitly represented as biomolecules undergoing reactions in dilute solution---there are no concealed degrees of freedom. As a result, the complexity of implementing minimal systems that exhibit efficient measurement and feedback is made clear, and ambiguities are eliminated.  \teo{In particular, we have outlined a molecular mechanism for implementing sequential measurement and feedback in an explicit setting,}
providing clarity not only to the extended correlation-exploiting devices, but also to our representation of the canonical Szilard engine itself. The continuing debate surrounding such devices (see references in \cite{ouldridge2018importance,parrondo2015thermodynamics}) shows the importance of a concrete physical representation.

For an environment with no structure---without correlations between successively encountered molecular bits---there is no need to process information and all of the available free energy can be extracted as work without use of a memory or any decision making. For a Markovian array with non-zero correlations between consecutive bits, we show that a simple two-state memory that can select one of two work extraction protocols can extract all of the stored free energy in the environment. The two-state memory is sufficient to carry all of the available information about the future of the environment forward, and we have identified a protocol that is simultaneously optimal for updating the memory according to the current input, and exploiting said input. For a more complex environment, involving a hidden variable that can only be inferred by the machine through noisy measurements, we argue that a machine with a finite memory cannot extract all of the available free energy as work. We demonstrate that in such a setting, a more complex strategy involving effectively averaging over a batch of molecules can be advantageous if correlations are long-ranged, but noise is substantial. This is similar to the result in \cite{seoane2018information} that a more complex predictive model is advantageous in a more complex environment, but, in this paper, we give an explicit physical model for how our machines measure and exploit the environment. In our design, the complexity of the mechanism involved the ability to couple to multiple inputs simultaneously; we predict that alternatives (such as systems with larger memories and more possible decision) would also show the potential for improved performance. 

A real living system must not only extract enough resources from its environment on average, but also over short intervals. In any fluctuating environment, an unlucky sequence of events might lead to starvation and death. We probe this situation in our minimal setting by considering the typical infimum (lowest point) of the work extracted by our devices, which represents the typical scale of negative fluctuations. We find that the more sophisticated inference strategy considered here also has smaller negative fluctuations when it is favourable on average, suggesting that it truly can be advantageous. In a minimal living system, reduced negative fluctuations would correspond to the need for a smaller reserve of energy, and the ability to produce viable offspring more quickly, since each offspring would need to be provided with the reserves to deal with typical negative fluctuations for a large fraction to survive.      

The minimal devices we consider are clearly unnatural, and constitute only a first step towards understanding the physics of living or life-like systems that make simple decisions. \teo{A key feature of our design is that only single copies of some molecular species are present; it is an open question how to design optimal systems in which the information-processing components, and indeed the inputs, fluctuate more widely. We do note, however, that much of the molecular decision making within cells occurs at the level of transcription factors binding to DNA---these transcription factor binding sites are present with a low and predictable copy number. } Future work will focus on constructing minimal models in which the systems are autonomous, requiring no external control, and power their own information-processing tasks by the free energy harvested. \teo{A major challenge here would be to implement reliable and efficient measurement and feedback without an externally imposed clock that allows sequential operation, as in this work.}  More realistic environments of fluctuating chemical concentrations, \teo{rather than input molecules that arrive one-by-one}, will need to be considered. \teo{In such descriptions, it will be necessary to construct a more detailed kinetic model of the underlying elementary reactions than we present here.} A deeper question is whether we can design concrete systems  that actually learn the statistics of their environment, evolving the parameters of their decision making process towards an optimal strategy, rather than having fixed, optimal parameters as in this work. 

Despite the simplicity of our current approach, however, we believe that concrete lessons can be drawn for the physics of living or life-like systems making simple decisions. In our physical model, successively more complex, and potentially costly, information-processing architectures perform better in successively more complex environments. We would expect that the information-processing carried out by living organisms reflects a similar trade-off: more complex decision-making strategies are more worthwhile in environments that exhibit statistical structure over time scales that are long compared to the decision-making time, and large fluctuations that must not be misinterpreted. We also expect that true evolved strategies will not optimise exploitation of the environment on average in isolation; strategies should also be designed to hedge against the risk of negative short-term fluctuations, to a degree that depends upon the cost of storing resources that compensate for these fluctuations.  

\section{Data availability}
The code and data to produce the figures in this paper can be found at \url{https://doi.org/10.5281/zenodo.1976932}.

\section{Acknowledgements}
T. E. O. acknowledges support from a Royal Society University Research Fellowship and R. A. B. acknowledges support from an Imperial College London AMMP studentship.

\bibliography{bibliography2}
\bibliographystyle{unsrt}

\clearpage

\begin{appendix}
\section{The molecular hook}
\label{sec:apphook}
The input molecules can be moved reversibly between their boxes and the reaction volume. Hooks to reversibly move molecules between different volumes has been previously discussed in the supplementary material of \cite{ouldridge2017fundamental}. First the input molecule must be attached to the hook. This can be done, for example, using the chemical reactions
\begin{align}
	\text{H}_\text{B}+\x+\fs[5]&\rightleftharpoons\text{H}_\text{B}\x+\f[5],\nonumber\\
	\text{H}_\text{B}+\xs+\fs[5]&\rightleftharpoons\text{H}_\text{B}\xs+\f[5],
	\label{eq:hookattach}
\end{align}
where H$_\text{B}$ is the hook with no molecule attached inside the input molecule box and H$_\text{B}\x$ and H$_\text{B}\xs$ are the hook with an input molecule attached. The binding of the hook is insensitive to the state of the input molecule.

The $\x$/$\xs$ molecule can be transferred to the reaction volume in the following steps. Initially, $[\f[5]]=[\fs[5]]=0$. After the hook is introduced to the box $[\f[5]]$ is increased up to a value of $f_5$. Then, $[\fs[5]]$ is increased to a value $f_5^*\gg f_5$. Now the $\x$/$\xs$ molecule is for certain attached to the hook. Next, the hook is transferred from the input box to the reaction volume. This transfer can be done either chemically using a conformational change in the hook molecule mediated by another pair of fuel molecules as in the equations:
\begin{align}
	\text{H}_\text{B}\x+\fs[6]&\rightleftharpoons\text{H}_\text{RV}\x+\f[6],\nonumber\\
	\text{H}_\text{B}\xs+\fs[6]&\rightleftharpoons\text{H}_\text{RV}\xs+\f[6],
	\label{eq:hookattach2}
\end{align}
or, alternatively, by a purely mechanical quasistatic process.

Then, to release the input molecule from the hook inside the reaction volume the reverse of the protocol the attach the input molecule is used. Initially $[\f[5]]=f_5$ and $[\fs[5]]=f_5^*$ with $f_5^*\gg f_5$. Then, $[\fs[5]]$ is decreased to zero and, subsequently, $[\f[5]]$ is decreased to zero. Now the input molecule is for certain released from the hook inside the reaction volume.

Subsequently, the measurement and extraction protocols can be carried out. Afterwards, the input molecule, which could have changed state, is moved back to its box. This is done by attaching the input molecule to the hook, moving the hook and detaching the input molecule from the hook in the exact reverse of the protocol to transfer the input molecule to the reaction volume.

The free energy of the input molecule is the same whether it is in the $\x$ or $\xs$ state. In general it is not the case that the free energy of the input molecule when it is in its box, $\mathcal{F}_\text{box}$, is the same as the free energy of the input when it is in the reaction volume, $\mathcal{F}_\text{reaction volume}$. Therefore, if the input molecule is transferred reversibly from its box to the reaction volume then the control must have done a work of $\mathcal{F}_\text{reaction volume}-\mathcal{F}_\text{box}$ on the input molecule. If the molecule is moved back to the box reversibly then this work is recovered.

Clearly, this principle can be extended to moving multiple molecules into the reaction volume as required by the batch machines, and then back again to the reaction volumes. Again, the whole process requires no net work. The initial transfer is not reversible for a structured environment, however, since the order of molecules is randomised within the reaction volumes.

\section{Work calculation for Biochemical Szilard engine}
\subsection{Measurement}
\label{sec:appszilardmeasure}
The measurement follows the optimal copy protocol in \cite{ouldridge2017thermodynamics}. Initially, the memory molecule is in states $\m[0]$ and $\m[1]$ with equal probability and the input molecule is independently in states $\x$ and $\xs$ with equal probability. The state of the input is measured using the equations
\begin{align}
	\m[0]+\x+\fs[1]&\rightleftharpoons\m[1]+\x+\f[1],\nonumber\\
	\m[0]+\xs+\fs[2]&\rightleftharpoons\m[1]+\xs+\f[2].
	\label{eq:appendixmeasure}
\end{align}
The free energy difference $\g[1]=\mu_{\f[1]}-\mu_{\fs[1]}=\g[1]^0+\ln\frac{[\f[1]]}{[\fs[1]]}$, where $\g[1]^0$ depends on the intrinsic nature of the $\f[1]$ and $\fs[1]$ molecules and the reaction volume but not their concentrations, is quasistatically changed from 0 to $\infty$ and $\g[2]=\mu_{\f[2]}-\mu_{\sf[2]}=\g[2]^0+\ln\frac{[\f[2]]}{[\fs[2]]}$ is quasistatically changed from 0 to $-\infty$.

Firstly, let us assume that the input molecule is in state $\x$, which occurs with prob of $1/2$. In this case, only the first reaction in equation \ref{eq:appendixmeasure} can occur. At any point in the process there is a probability $p(\m[1])$ that the memory molecule is in state $\m[1]$. This probability only changes with a corresponding change in the number of $\f[1]$ and $\fs[1]$ in the buffer. If $p(\m[1])$ changes by a small amount $\text{d}p(\m[1])$ then $\text{d}p(\m[1])$ $\fs[1]$ are converted into $\f[1]$ so a work of $\text{d}p(\m[1])\g[1]$ is done on the buffer connected to the reaction volume. Therefore, in a process a work is done on the buffers of
\begin{equation}
    W=\int_{t_i}^{t_f}\text{d}t\dir{p(\m[1])}{t}\g[1](t).
    \label{eq:appendixtimeintegral}
\end{equation}

Because the change in concentration of the fuels is quasistatic, at all times in the process the memory molecule is in equilibrium with the fuel buffer the reaction volume is connected to. Therefore,
\begin{equation}
    p(\m[1])=\frac{\e{-\beta\g[1]}}{1+\e{-\beta\g[1]}},
    \label{eq:appendixpm1}
\end{equation}
where $\beta=\frac{1}{\kb T}$.

The fact that only dependence $p(\m[1])$ has on time is through $\g[1]$ means that equation \ref{eq:appendixtimeintegral} can be converted into an integral over $\g[1]$ instead. Because the change is quasistatic the particular function of time that $\g[1]$ is does not matter. Only the change in $\g[1]$ matters. Therefore,
\begin{equation}
    W=\int_0^\infty\text{d}\g[1]\dir{p(\m[1])}{\g[1]}\g[1].
\end{equation}
Now to get the work we simply have to use equation \ref{eq:appendixpm1} and evaluate the integral. It is convenient to first integrate by parts to get
\begin{equation}
    W=\left[p(\m[1])\g[1]\right]_0^\infty-\int_0^\infty\text{d}\g[1]p(\m[1]),
\end{equation}
and then exploit equation \ref{eq:appendixpm1} to get
\begin{align}
    W&=\left[\frac{\e{-\beta\g[1]}}{1+\e{-\beta\g[1]}}\g[1]\right]_0^\infty-\int_0^\infty\text{d}\g[1]\frac{\e{-\beta\g[1]}}{1+\e{-\beta\g[1]}}\nonumber\\
    &=-\left[-\frac{1}{\beta}\ln\left(1+\e{-\beta\g[1]}\right)\right]_0^\infty\nonumber\\
    &=-\kb T\ln2,
    \label{eq:appendixmeasureintegral1}
\end{align}
using l'H\^opital's rule for the $\g[1]\rightarrow\infty$ limit in the first line. A negative work corresponds to a decrease in free energy of the buffers.

Alternatively, there is a probability of $1/2$ that the input molecule is $\xs$ so only the second reaction in equation \ref{eq:appendixmeasure} can occur. In this case
\begin{equation}
    p(\m[1])=\frac{\e{-\beta\g[2]}}{1+\e{-\beta\g[2]}}.
    \label{eq:appendixpm12}
\end{equation}
where $\g[2]=\mu_{\f[2]}-\mu_{\fs[2]}=\g[2]^0+\ln\frac{[\f[2]]}{[\fs[2]]}$ and the work done on the buffers is
\begin{align}
    W&=\int_0^{-\infty}\text{d}\g[2]\dir{p(\m[1])}{\g[2]}\g[2]\nonumber\\
    &=\left[\frac{\e{-\beta\g[2]}}{1+\e{-\beta\g[2]}}\g[2]\right]_0^{-\infty}-\int_0^{-\infty}\text{d}\g[2]\frac{\e{-\beta\g[2]}}{1+\e{-\beta\g[2]}}\nonumber\\
    &=\left[\frac{\e{-\beta\g[2]}}{1+\e{-\beta\g[2]}}\g[2]+\frac{1}{\beta}\ln\left(1+\e{-\beta\g[2]}\right)\right]_0^{-\infty}.
    \label{eq:appendixmeasureintegral2}
\end{align}
To evaluate the upper limit it is convenient to substitute in equation \ref{eq:appendixpm12}
\begin{align}
    W&=\Big[-p(\m[1])\ln p(\m[1])-\big(1-p(\m[1])\big)\ln\big(1-p(\m[1])\big)\Big]_{1/2}^1\nonumber\\
    &=-\kb T\ln2.
    \label{eq:appendixmeasureintegral2transformation}
\end{align}

Each of these possibilities is equally likely so the expected work is
\begin{equation}
    W=-\kb T\ln2.
\end{equation}

\subsection{Extraction}
\label{sec:appszilardextract}
Now the system is either in state $(\x,\m[0])$ or $(\xs,\m[1])$. Work is extracted from this high free energy state using the reactions
\begin{align}
	\m[0]+\x+\fs[3]&\rightleftharpoons\m[0]+\xs+\f[3],\nonumber\\
	\m[1]+\x+\fs[4]&\rightleftharpoons\m[1]+\xs+\f[4].
    \label{eq:appendixszilardextract}
\end{align}

If the system is in state $(\x,\m[0])$ then only the first reaction in equation \ref{eq:appendixszilardextract} can occur. In this case the probability of the input molecule being in the $\xs$ state is:
\begin{equation}
    p(\xs)=\frac{\e{-\beta\g[3]}}{1+\e{-\beta\g[3]}},
\end{equation}
where $\g[3]=\mu_{\fs[3]}-\mu_{\f[3]}=\g[3]^0+\ln\frac{[\fs[3]]}{[\f[3]]}$. As $\g[3]$ is changed from $\infty$ to 0, we obtain
\begin{align}
    W&=\int_{\infty}^0\text{d}\g[3]\dir{p(\xs)}{\g[3]}\g[3]\nonumber\\
    &=\left[\frac{\e{-\beta\g[3]}}{1+\e{-\beta\g[3]}}\g[3]\right]_{\infty}^0-\int_{\infty}^0\text{d}\g[3]\frac{\e{-\beta\g[3]}}{1+\e{-\beta\g[3]}}\nonumber\\
    &=-\left[-\frac{1}{\beta}\ln\left(1+\e{-\beta\g[3]}\right)\right]_{\infty}^0\nonumber\\
    &=\kb T\ln2.
\end{align}
This is exactly the same calculation as equation \ref{eq:appendixmeasureintegral1}. The sign is positive because the free energy of the fuel molecule buffers is now increased.

If the system is in state $(\xs,\m[1])$ then only the second reaction in equation \ref{eq:appendixszilardextract} can occur.  In this case the probability of the input molecule being in the $\xs$ state is:
\begin{equation}
    p(\xs)=\frac{\e{-\beta\g[4]}}{1+\e{-\beta\g[4]}},
\end{equation}
where $\g[4]=\mu_{\fs[4]}-\mu_{\f[4]}=\g[4]^0+\ln\frac{[\fs[4]]}{[\f[4]]}$. As $\g[4]$ is changed from $-\infty$ to 0, we obtain
\begin{align}
    W&=\int_{-\infty}^0\text{d}\g[4]\dir{p(\xs)}{\g[4]}\g[4]\nonumber\\
    &=\left[\frac{\e{-\beta\g[4]}}{1+\e{-\beta\g[4]}}\g[4]\right]_{-\infty}^0-\int_{-\infty}^0\text{d}\g[4]\frac{\e{-\beta\g[4]}}{1+\e{-\beta\g[4]}}\nonumber\\
    &=\left[\frac{\e{-\beta\g[4]}}{1+\e{-\beta\g[4]}}\g[4]+\frac{1}{\beta}\ln\left(1+\e{-\beta\g[4]}\right)\right]_{-\infty}^0\nonumber\\
    &=\kb T\ln2.
\end{align}
This is exactly the same calculation as equations \ref{eq:appendixmeasureintegral2} and \ref{eq:appendixmeasureintegral2transformation}.

Each of these possibilities is equally likely so the expected work is
\begin{equation}
    W=\kb T\ln2.
\end{equation}
Therefore, in a measure and extract cycle the net work done by the fuel molecule buffers is zero.

\section{Quasistatic embeddability of Markov machine}
\label{app:embed}
Any transformation of a probability distribution over discrete states can be represented by a stochastic matrix. A quasistatic embedding is a non-homogeneous continuous time Markov chain that produces such a transformation with no entropy production \cite{owen2017number}. It is not possible to find such an embedding for all stochastic matrices. For some stochastic matrices the state-space must be extended with additional `hidden' states before a quasistatic embedding can be found. Owen {\it et al.} \cite{owen2017number} have found bounds on the number of hidden state required.

We can apply the results of \cite{owen2017number} to the Markov machine. The joint system of the input molecule and memory molecule has four states. We order them $(\x\m[0],\x\m[1],\xs\m[0],\xs\m[1])$. The transformation that measures the state of the input molecule to the state of the memory molecule and takes the input molecule to its equilibrium distribution is then
\begin{equation}
    P=\begin{pmatrix}
    1/2 & 1/2 & 0 & 0 \\
    0 & 0 & 1/2 & 1/2 \\
    1/2 & 1/2 & 0 & 0 \\
    0 & 0 & 1/2 & 1/2
    \end{pmatrix}.
    \label{eq:transitionmatrix}
\end{equation}

The determinant of $P$ is zero so according to \cite{owen2017number} the lower bound on the number of additional hidden states required for a quasistatic embedding is zero.

The upper found on the number of hidden states required is $r^+(P)-1$ where $r^+(P)$ is the nonnegative rank of $P$. For an $n\times n$ stochastic matrix, $M$, the nonnegative rank is the smallest $m$ such that $M$ can be written $M=RS$ where $R$ is a $n\times m$ stochastic matrix and $S$ is a $m\times n$ stochastic matrix.

$P$ can be written as
\begin{equation}
	P=\begin{pmatrix}
		\frac{1}{2} & 0 & \frac{1}{2} & 0 \\[2pt]
        \frac{1}{2} & 0 & \frac{1}{2} & 0 \\
        0 & \frac{1}{2} & 0 & \frac{1}{2} \\[2pt]
        0 & \frac{1}{2} & 0 & \frac{1}{2}
	\end{pmatrix}
    =
    \begin{pmatrix}
		0 & \frac{1}{2} \\[2pt]
        0 & \frac{1}{2} \\
        \frac{1}{2} & 0 \\[2pt]
        \frac{1}{2} & 0
	\end{pmatrix}
    \begin{pmatrix}
    	0 & 1 & 0 & 1 \\
        1 & 0 & 1 & 0
    \end{pmatrix},
\end{equation}
so the nonnegative rank of this matrix is two and the upper bound on the number of additional hidden states required is one.

In the main text we see that the actual number of hidden states required is zero because we construct the protocol explicitly.

\section{Work calculation for Markov machine}
\subsection{Measurement}
\label{sec:appmarkovmeasure}
The measurement follows the optimal copy protocol in \cite{ouldridge2017thermodynamics}. Instead of the input and memory molecules initially being independent, as they were for the Szilard engine, the molecules now start off correlated.

The reactions
\begin{align}
	\m[0]+\x+\fs[1]&\rightleftharpoons\m[1]+\x+\f[1],\nonumber\\
	\m[0]+\xs+\fs[2]&\rightleftharpoons\m[1]+\xs+\f[2].
	\label{eq:appendixmarkovmeasure}
\end{align}
are, again, used. If the input is $\x$ then only the first reaction in equation \ref{eq:appendixmarkovmeasure} can occur. Now $p(\m[1]|\x)$ is not $1/2$, it is set by the input process. To set the memory molecule to $\m[0]$ using the least work $\g[1]=\mu_{\f[1]}-\mu_{\fs[1]}=\g[1]^0+\ln\frac{[\f[1]]}{[\fs[1]]}$ must be initially set to $\g[1]=\g[\text{offset}]$ such that
\begin{equation}
    p(\m[1]|\x)=\frac{\e{-\beta\g[\text{offset}]}}{1+\e{-\beta\g[\text{offset}]}},
\end{equation}
and then quasistatically changed to $\g[1]=\infty$. Therefore, the work done is
\begin{align}
    W&=\int_{\g[\text{offset}]}^\infty\text{d}\g[1]\dir{p(\m[1])}{\g[1]}\g[1]\nonumber\\
    &=\left[\frac{\e{-\beta\g[1]}}{1+\e{-\beta\g[1]}}\g[1]\right]_{\g[\text{offset}]}^\infty-\int_{\g[\text{offset}]}^\infty\text{d}\g[1]\frac{\e{-\beta\g[1]}}{1+\e{-\beta\g[1]}}\nonumber\\
    &=-\frac{\e{-\beta\g[\text{offset}]}}{1+\e{-\beta\g[\text{offset}]}}\g[\text{offset}]-\left[-\frac{1}{\beta}\ln\left(1+\e{-\beta\g[1]}\right)\right]_{\g[\text{offset}]}^\infty\nonumber\\
    &=-\frac{\e{-\beta\g[\text{offset}]}}{1+\e{-\beta\g[\text{offset}]}}\g[\text{offset}]-\frac{1}{\beta}\ln\left(1+\e{-\beta\g[\text{offset}]}\right)\nonumber\\
    &=\frac{1}{\beta}\Big(p(\m[1]|\x)\ln p(\m[1]|\x)+\big(1-p(\m[1]|\x)\big)\ln\big(1-p(\m[1]|\x)\big)\Big).
\end{align}

Similarly, if the input molecule is $\xs$ then $\g[2]=\mu_{\f[2]}-\mu_{\fs[2]}=\g[2]^0+\ln\frac{[\f[2]]}{[\fs[2]]}$ must be initially set to$\g[2]=-\g[\text{offset}]$ such that
\begin{equation}
    p(\m[1]|\xs)=\frac{\e{\beta\g[\text{offset}]}}{1+\e{\beta\g[\text{offset}]}},
\end{equation}
and then quasistatically changed to $\g[1]=-\infty$. Therefore, the work done is
\begin{align}
    W&=\int_{-\g[\text{offset}]}^{-\infty}\text{d}\g[2]\dir{p(\m[1])}{\g[2]}\g[2]\nonumber\\
    &=\left[\frac{\e{-\beta\g[2]}}{1+\e{-\beta\g[2]}}\g[2]\right]_{-\g[\text{offset}]}^{-\infty}-\int_{-\g[\text{offset}]}^{-\infty}\text{d}\g[2]\frac{\e{-\beta\g[2]}}{1+\e{-\beta\g[2]}}\nonumber\\
    &=\left[\frac{\e{-\beta\g[2]}}{1+\e{-\beta\g[2]}}\g[2]+\frac{1}{\beta}\ln\left(1+\e{-\beta\g[2]}\right)\right]_{-\g[\text{offset}]}^{-\infty}\nonumber\\
    &=\frac{\e{\beta\g[\text{offset}]}}{1+\e{\beta\g[\text{offset}]}}\g[\text{offset}]-\frac{1}{\beta}\ln\left(1+\e{\beta\g[\text{offset}]}\right)\nonumber\\
    &=\frac{1}{\beta}\Big(p(\m[1]|\xs)\ln p(\m[1]|\xs)\nonumber\\
    &+\big(1-p(\m[1]|\xs)\big)\ln\big(1-p(\m[1]|\xs)\big)\Big).
\end{align}

The first case occurs with probability $p(\x)$ and the second occurs with probability $p(\xs)$ so the expected work is
\begin{align}
    \begin{split}
        W=&p(\x)\frac{1}{\beta}\Big(p(\m[1]|\x)\ln p(\m[1]|\x)\\
        &+\big(1-p(\m[1]|\x)\big)\ln\big(1-p(\m[1]|\x)\big)\Big)\\
        &+p(\xs)\frac{1}{\beta}\Big(p(\m[1]|\xs)\ln p(\m[1]|\xs)\\
        &+\big(1-p(\m[1]|\xs)\big)\ln\big(1-p(\m[1]|\xs)\big)\Big)
    \end{split}\nonumber\\
    =&-\frac{1}{\beta}H[M_i|X_{i+1}].
\end{align}

The random variable $M_i$ is an exact copy of $X_i$ so
\begin{align}
    W&=-\frac{1}{\beta}H[M_i|X_{i+1}]\nonumber\\
    &=-\frac{1}{\beta}H[X_i|X_{i+1}].
\end{align}
The input process is stationary so $H[X_i]=H[X_{i+1}]$ so \cite{Elements_of_Information_Theory}
\begin{align}
    W&=-\frac{1}{\beta}H[X_i|X_{i+1}]\nonumber\\
    &=-\frac{1}{\beta}\Big(H[X_{i+1}|X_i]+H[X_i]-H[X_{i+1}]\Big)\nonumber\\
    &=-\frac{1}{\beta}H[X_{i+1}|X_i].
\end{align}

$H[X_{i+1}|X_i]\leq\ln2$ so the fact that the memory molecule and input molecule are initially correlated means that the measurement requires less work to be done on the system by the fuel molecule buffers.

\subsection{Extraction}
\label{sec:appmarkovextract}
The extraction process is exactly the same as for the biochemical Szilard engine. Therefore, the work done on the fuel molecule buffers is $\kb T\ln2$ so the net work per input molecule is
\begin{equation}
    W=\kb T\Big(\ln2-H[X_{i+1}|X_i]\Big).
\end{equation}

\section{DNA design of Biochemical Szilard engine and Markov machine}
\label{sec:appszilarddna}
\begin{figure}[h]
	\includegraphics[width=1\linewidth]{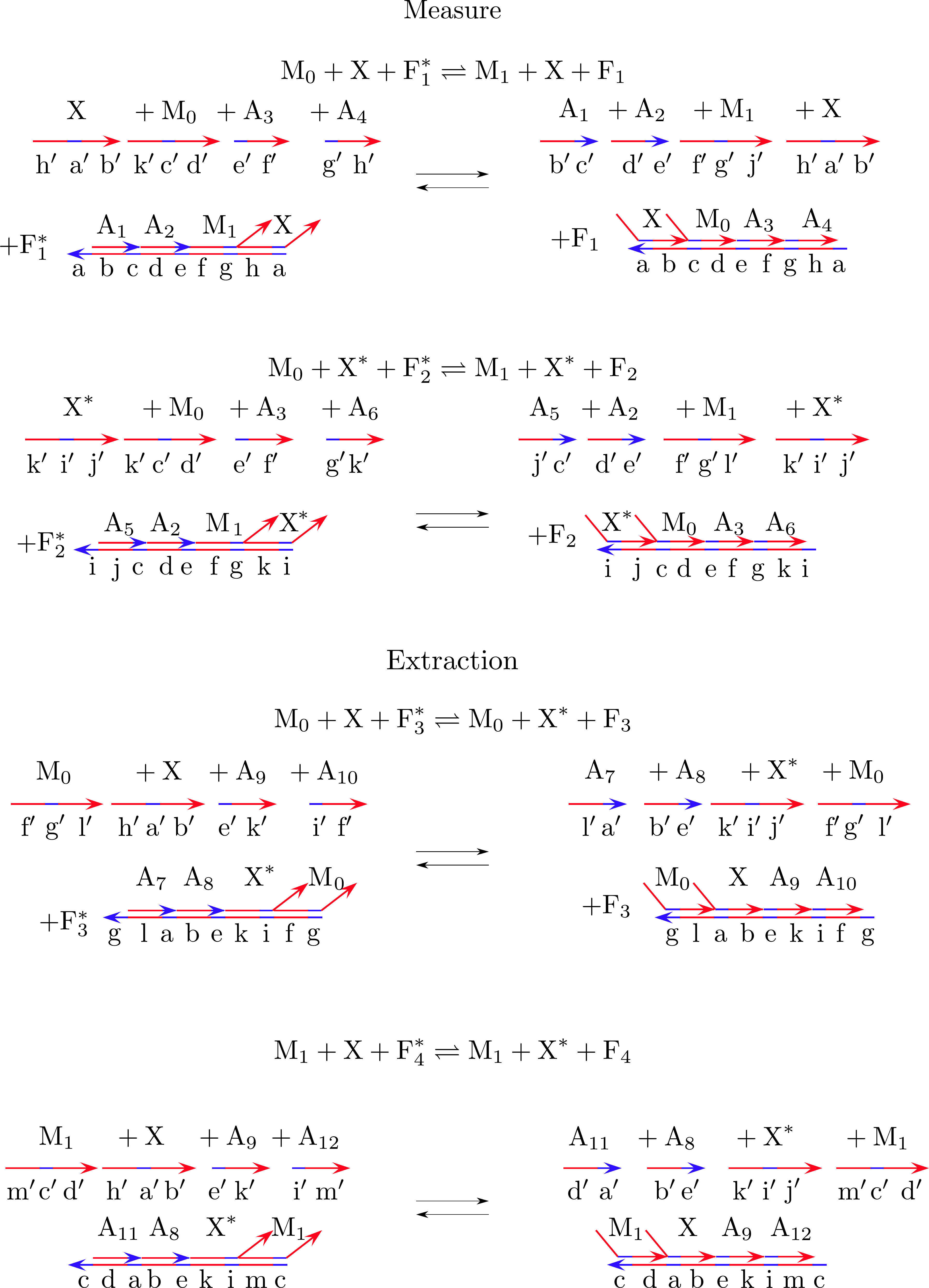}
    \caption{\emph{A domain level DNA-based design to implement the measurement and work extraction reactions of the Biochemical Szilard engine.} The arrowheads show the $3'$ end of the strands. The label `a' represents a sequence of bases (domain) and `a$'$' represents the complimentary sequence.}
    \label{fig:szilarddna}
\end{figure}

In this section we present a domain level DNA-based design to implement the measurement and work extraction reactions of the Biochemical Szilard engine and Markov machine using DNA strand displacement. The design is shown in figure \ref{fig:szilarddna}. Our designs leverage the general construction of \cite{soloveichik2010dna}.

The nature of DNA strand displacement reactions means that additional auxiliary strands, labelled A$_1$ to A$_{12}$, are required. We assume that these strands are always present in the reaction volume in excess.

\section{Work calculation for Binary batch machine}
\subsection{Measurement}
\label{sec:appbinarybatchmeasure}
The measurement follows the optimal copy protocol in \cite{ouldridge2017thermodynamics}. The measurement is done using the reactions
\begin{align}
	\m[0]+\frac{N+1}{2}\x+\fs[1]&\rightleftharpoons\m[1]+\frac{N+1}{2}\x+\f[1], \nonumber\\
    \m[0]+\frac{N+1}{2}\xs+\fs[2]&\rightleftharpoons\m[1]+\frac{N+1}{2}\xs+\f[2],
    \label{eq:appendixbatchmeasureodd}
\end{align}
when $N$ is odd and
\begin{align}
	\m[0]+\frac{N}{2}\x+\fs[1]&\rightleftharpoons\m[1]+\frac{N}{2}\x+\f[1],\nonumber\\
    \m[0]+\left(\frac{N}{2}+1\right)\xs+\fs[2]&\rightleftharpoons\m[1]+\left(\frac{N}{2}+1\right)\xs+\f[2],
    \label{eq:appendixbatchmeasureeven}
\end{align}
when $N$ is even.

Let $J_i$ be the random variable representing whether  the number of $\xs$ molecules in batch $i$ is greater than $N/2$ ($J_i=1$ if true, 0 otherwise). The measurement process is exactly the same as for the Markov machine except that the chemical potential differences, $\g[1]=\mu_{\f[1]}-\mu_{\fs[1]}=\g[1]^0+\ln\frac{[\f[1]]}{[\fs[1]]}$ and $\g[2]=\mu_{\f[2]}-\mu_{\fs[2]}=\g[2]^0+\ln\frac{[\f[2]]}{[\fs[2]]}$, are initially set to $\g[1]=\g[\text{offset}]^1$ and $\g[2]=\g[\text{offset}]^2$ such that
\begin{equation}
    p(\m[1]|J_i=0)=\frac{\e{-\beta\g[\text{offset}]^1}}{1+\e{-\beta\g[\text{offset}]^1}},
\end{equation}
and
\begin{equation}
    p(\m[1]|J_i=1)=\frac{\e{-\beta\g[\text{offset}]^2}}{1+\e{-\beta\g[\text{offset}]^2}}.
\end{equation}
Therefore, the work done is
\begin{equation}
    W=-\kb TH[M_i|J_{i+1}],
\end{equation}
and, similar to the Markov machine, $M_i$ is an exact copy of $J_i$ and $H[J_{i+1}]=H[J_i]$ so
\begin{equation}
    W=-\kb TH[J_{i+1}|J_i].
\end{equation}

\subsection{Extraction}
\label{sec:appbinarybatchextract}
The extraction step uses the same reactions as the biochemical Szilard engine and the Markov machine:
\begin{align}
	\m[0]+\x+\fs[3]&\rightleftharpoons\m[0]+\xs+\f[3],\nonumber\\
	\m[1]+\x+\fs[4]&\rightleftharpoons\m[1]+\xs+\f[4].
    \label{eq:appendixbinarybatchextract}
\end{align}
However, the protocol of the chemical potential differences must be different. In the biochemical Szilard engine and Markov machine, if the memory molecule was in the state $\m[0]$ then the input molecule would be for certain in the state $\x$. However, in the binary batch machine if the memory molecule is in the state $\m[0]$ then there is a nonzero probability that some of the input molecules in the batch are in state $\xs$.

The chemical potential differences, $\g[3]=\mu_{\fs[3]}-\mu_{\f[3]}=\g[3]^0+\ln\frac{[\fs[3]]}{[\f[3]]}$ and $\g[4]=\mu_{\fs[4]}-\mu_{\f[4]}=\g[4]^0+\ln\frac{[\fs[4]]}{[\f[4]]}$, are started at $\g[3]=\g[\text{offset}]^3$ and $\g[4]=\g[\text{offset}]^4$ where $\g[\text{offset}]^3$ and $\g[\text{offset}]^4$ are set so that
\begin{equation}
    \hat{p}_0=\frac{\e{-\beta\g[\text{offset}]^3}}{1+\e{-\beta\g[\text{offset}]^3}},
\end{equation}
and
\begin{equation}
    \hat{p}_1=\frac{\e{-\beta\g[\text{offset}]^4}}{1+\e{-\beta\g[\text{offset}]^4}},
\end{equation}
First, there is an irreversible relaxation in the batch from the initial input distribution, which depends on the input stochastic process, to a binomial distribution over the number of $\xs$ molecules with a mean of $N\hat{p}_0$ or $N\hat{p}_1$. If the memory molecule is is $\m[0]$, the work extracted in this relaxation is
\begin{equation}
    W_\text{relax}=\g[\text{offset}]^3\left(N\hat{p}_0-\langle\xs\rangle_{\m[0]}^\text{initial}\right),
\end{equation}
where $\langle\xs\rangle_{\m[0]}^\text{initial}$ is the expected number of $\xs$ in the batch initially. If the memory molecule is is $\m[1]$, the work extracted in this relaxation is
\begin{equation}
    W_\text{relax}=\g[\text{offset}]^4\left(N\hat{p}_1-\langle\xs\rangle_{\m[1]}^\text{initial}\right),
\end{equation}
where $\langle\xs\rangle_{\m[1]}^\text{initial}$ is the expected number of $\xs$ in the batch initially. Then, $\g[3]$ and $\g[4]$ are quasistatically changed to zero. If the memory molecule is in state $\m[0]$ the work that is done in this quasistatic step is
\begin{align}
    \hspace{-0.3cm}W_\text{q}=&N\int_{\g[\text{offset}]^3}^0\text{d}\g[3]\dir{p(\xs)}{\g[3]}\g[3]\nonumber\\
    =&N\left(\left[\frac{\e{-\beta\g[3]}}{1+\e{-\beta\g[3]}}\g[3]\right]_{\g[\text{offset}]^3}^0-\int_{\g[\text{offset}]^3}^0\text{d}\g[3]\frac{\e{-\beta\g[3]}}{1+\e{-\beta\g[3]}}\right)\nonumber\\
    =&N\left[\frac{\e{-\beta\g[3]}}{1+\e{-\beta\g[3]}}\g[3]+\frac{1}{\beta}\ln\left(1+\e{-\beta\g[3]}\right)\right]_{\g[\text{offset}]^3}^0\nonumber\\
    =&N\left(\frac{1}{\beta}\ln2-\frac{\e{-\beta\g[\text{offset}]^3}}{1+\e{-\beta\g[\text{offset}]^3}}\g[\text{offset}]^3-\frac{1}{\beta}\ln\left(1+\e{-\beta\g[\text{offset}]^3}\right)\right)\nonumber\\
    =&N\frac{1}{\beta}\Big(\ln2+\hat{p}_0\ln \hat{p}_0+(1-\hat{p}_0)\ln(1-\hat{p}_0)\Big).
\end{align}
Similarly, if the memory molecule is in state $\m[1]$ the work that is done in this quasistatic step is
\begin{equation}
    W_\text{q}=N\frac{1}{\beta}\Big(\ln2+\hat{p}_1\ln \hat{p}_1+(1-\hat{p}_1)\ln(1-\hat{p}_1)\Big).
\end{equation}
Therefore, if the memory molecule is in state $\m[0]$ the total work that is done in the irreversible relaxation and quasistatic steps is
\begin{equation}
    N\frac{1}{\beta}\Big(\ln2+\langle\xs\rangle_{\m[0]}^\text{initial}/N\ln \hat{p}_0+(1-\langle\xs\rangle_{\m[0]}^\text{initial}/N)\ln(1-\hat{p}_0)\Big).
\end{equation}
This is maximised if $\hat{p}_0=\langle\xs\rangle_{\m[0]}^\text{initial}/N$. Similarly, if the memory molecule is in state $\m[0]$ the work is maximised by setting $\hat{p}_1=\langle\xs\rangle_{\m[1]}^\text{initial}/N$. i.e. $\hat{p}_0$ is the probability that an input molecule in the batch is $\xs$ if $J_i=0$ and $\hat{p}_1$ is the probability that an input molecule in the batch is $\xs$ if $J_i=1$. This means that initially no work is done on the fuel molecule buffers during the irreversible relaxation because on average there is no net change of number of $\xs$ in the batch.

Therefore, the expected work done in the extraction step is
\begin{align}
	W=\kb TN\Big(\ln2&+p(J_i=0)\big(\hat{p}_0\ln \hat{p}_0+(1-\hat{p}_0)\ln(1-\hat{p}_0)\big)\nonumber\\
    &+p(J_i=1)\big(\hat{p}_1\ln\hat{p_1}+(1-\hat{p}_1)\ln(1-\hat{p}_1)\big)\Big),
\end{align}
and the net work for the measure and extract cycle is
\begin{align}
	W=\kb TN\Big(\ln2&+p(J_i=0)\big(\hat{p}_0\ln \hat{p}_0+(1-\hat{p}_0)\ln(1-\hat{p}_0)\big)\nonumber\\
    &+p(J_i=1)\big(\hat{p}_1\ln\hat{p_1}+(1-\hat{p}_1)\ln(1-\hat{p}_1)\big)\Big)\nonumber\\
    -\kb TH[J_i|&J_{i-1}].
\end{align}

\section{DNA design of batch machine measurement}
\label{sec:appbatchdna}
\begin{figure*}[ht]
	\includegraphics[width=1\linewidth]{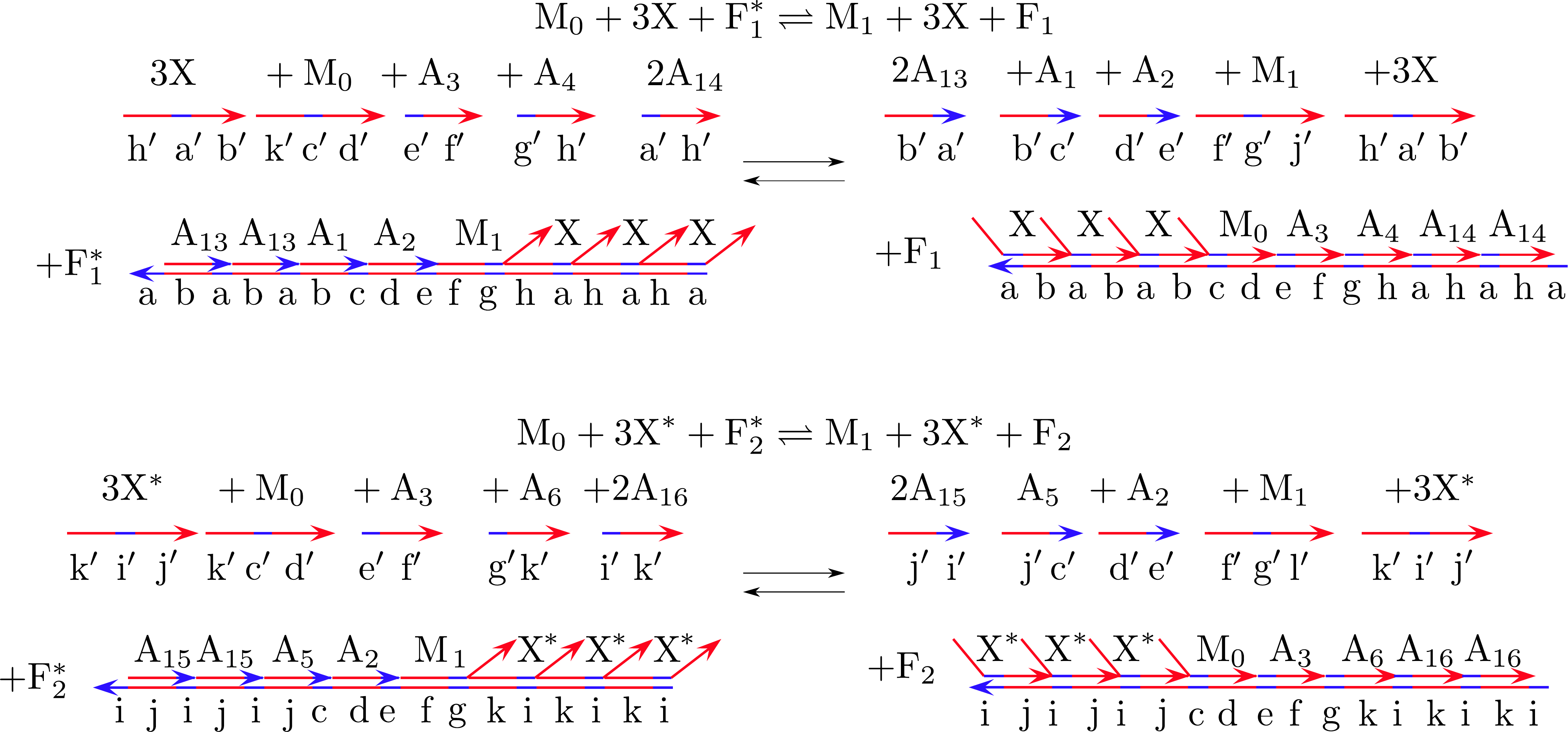}
    \caption{\emph{A domain level DNA-based design to implement the measurement reactions of the batch machines.} The arrowheads show the $3'$ end of the strands. The label `a' represents a sequence of bases (domain) and `a$'$' represents the complimentary sequence. The extraction reactions are the same as for the biochemical Szilard engine.}
    \label{fig:batch5dna}
\end{figure*}

In this section we present a domain level DNA-based design to implement the measurement reactions of the batch machine using DNA strand displacement. The design is shown in figure \ref{fig:batch5dna} for the case when $N=5$.  Our designs leverage the general construction of \cite{soloveichik2010dna}. This design is the same as the measurement reactions for the Biochemical Szilard engine and Markov machine except for that the gates are extended so that three $\x$ or $\xs$ strands must bind for the reaction to occur. In principle, the mechanism could be generalised to an arbitrary number of inputs---although this may prove challenging in practice. 

The nature of DNA strand displacement reactions means that additional auxiliary strands, labelled A$_1$ to A$_{16}$, are required. We assume that these strands are always present in the reaction volume in excess.

\section{Work calculation for full batch machine}
\label{sec:fullbatchwork}
\subsection{Measurement}
The measurement is exactly the same as for the binary batch machine so the work is
\begin{equation}
    W_\text{measure}=-\kb TH[J_{i+1}|J_i].
\end{equation}

\subsection{Extraction}
In this section we will not give an explicit chemical scheme to extract all of the work from an unordered batch of input molecules. We will simply calculate the available work. In equilibrium the number of $\xs$ molecules in the batch, $n(\xs)$, is described by a random variable $B^\text{eq}$, which is distributed as
\begin{equation}
	p\big(B^\text{eq}=n(\xs)\big)=\frac{1}{2^N}\frac{N!}{n(\xs)!(N-n(\xs))!}.
\end{equation}
If we define the free energy of each state of the unordered batch as
\begin{equation}
	\mathcal{F}(n(\xs))=-k_BT\ln p\big(B^\text{eq}=n(\xs)),
\end{equation}
then the equilibrium free energy is zero. Initially the number of $\xs$ molecules in the batch is described by a random variable $B^\text{initial}$ The free energy of the batch is initially
\begin{multline}
    \mathcal{F}_\text{B}(B^\text{initial})=\sum_{n(\xs)=0}^Np_\text{in}(n(\xs)\mathcal{F}(n(\xs))\\
    +\kb T\sum_{n(\xs)=0}^Np_\text{in}(n(\xs)\ln p_\text{in}(n(\xs),
\end{multline}
where $p_\text{in}(n(\xs)$ is the initial distribution over the number of $\xs$ in the batch and $B^\text{initial}$ is the random variable that describes the initial state of the batch. Therefore, using equation \ref{eq:informationisfreeenergy}, the free energy of the joint system of the batch and the memory molecule is
\begin{equation}
    \mathcal{F}_\text{joint}(B^\text{initial},M)=\mathcal{F}_\text{B}(B^\text{initial})+\mathcal{F}_\text{M}(M)+\kb T\mathcal{I}[B^\text{initial};M],
\end{equation}
where $M$ is the state of the memory molecule after the batch has been measured. After the work extraction the free energy of the joint system of the batch and the memory molecule is
\begin{equation}
    \mathcal{F}(B^\text{eq},M)=\mathcal{F}(B^\text{eq})+\mathcal{F}(M)+\kb T\mathcal{I}[B^\text{eq};M].
\end{equation}
As previously mention, $\mathcal{F}(B^\text{eq})=0$. It is also the case that $\mathcal{I}[B^\text{eq};M]=0$. Therefore, the maximum work that can be extracted is
\begin{align}
    W_\text{extract}&=\mathcal{F}_\text{joint}(B^\text{initial},M)-\mathcal{F}_\text{joint}(B^\text{eq},M)\nonumber\\
    &=\mathcal{F}_\text{B}(B^\text{initial})+\kb T\mathcal{I}[B^\text{initial};M]\nonumber\\
    &=\mathcal{F}_\text{B}(B^\text{initial})+\kb TH[M],
\end{align}
where the last line follows because knowing the state of $B^\text{initial}$gives you exact knowledge of the state of $M$ so $H[M|B^\text{initial}]=0$. Therefore, the net work extracted per batch is
\begin{equation}
    W=\mathcal{F}(B^\text{initial})+\kb T\mathcal{I}[J_{i+1},J_i].
\end{equation}

\end{appendix}

\end{document}